\documentclass[aps,prl,superscriptaddress,twocolumn,citeautoscript, longbibliography]{revtex4-1}
\usepackage{graphicx}
\usepackage{amsfonts}
\usepackage{amsmath}
\usepackage{amssymb}
\usepackage{bm}
\usepackage{textcomp}
\usepackage{verbatim}
\usepackage{xspace}
\usepackage{soul}
\usepackage{bbold}
\usepackage{mathtools}
\usepackage{dcolumn}
\usepackage{hyperref}
\usepackage[usenames]{color}
\usepackage{subfig}
\usepackage{physics}
\usepackage{units}
\usepackage[export]{adjustbox}
\usepackage{tablefootnote}
\usepackage[dvipsnames]{xcolor}
\usepackage{ragged2e}
\DeclareCaptionJustification{justified}{\justifying}
\def\beq{\begin{equation}}
\def\eeq{\end{equation}}

\def\fo{{F$\mathcal{O}$ \xspace}}

\def\afq{{AF$\mathcal{Q}$ \xspace}}
\def\afo{{AF$\mathcal{O}$ \xspace}}

\def\jeff{{$J_{\text{eff}}$ \xspace}}

\newcommand{\out}[1]{{}}

\newcommand{\bcnoo}{Ba$_2$Ca$_{1-\delta}$Na$_{\delta}$OsO$_6$ \xspace}

\begin{document}


\title{Polaron-driven switching of octupolar order in doped 5d$^2$ double perovskite}

\author{Dario Fiore Mosca}
\address{University of Vienna, Faculty of Physics, Center for Computational Materials Science, Kolingasse 14-16, 1090, Vienna, Austria}
\address{CPHT, CNRS, \'Ecole polytechnique, Institut Polytechnique de Paris, 91120 Palaiseau, France}
\address{Coll\`ege de France, Université PSL, 11 place Marcelin Berthelot, 75005 Paris, France}

\author{Lorenzo Celiberti}
\address{University of Vienna, Faculty of Physics, Center for Computational Materials Science, Kolingasse 14-16, 1090, Vienna, Austria}

\author{Leonid V. Pourovskii}
\address{CPHT, CNRS, \'Ecole polytechnique, Institut Polytechnique de Paris, 91120 Palaiseau, France}
\address{Coll\`ege de France, Université PSL, 11 place Marcelin Berthelot, 75005 Paris, France}

\author{Cesare Franchini}
\address{University of Vienna, Faculty of Physics, Center for Computational Materials Science, Kolingasse 14-16, 1090, Vienna, Austria}
\address{Department of Physics and Astronomy "Augusto Righi", Alma Mater Studiorum - Universit\`a di Bologna, Bologna, 40127 Italy}

\begin{abstract}

We investigate how doping-induced small polarons impact the low-temperature multipolar orders of the $5d^2$ double perovskite \(\mathrm{Ba_2CaOsO_6}\). By computing  intersite exchange interactions between \(5d^1\) localized hole polarons and \(5d^2\) magnetic ions from first principles, we demonstrate the reversal of the dominant octupolar exchange from ferromagnetic to antiferromagnetic. Solving the corresponding effective Hamiltonian we find this reversal  to 
account for the progressive suppression of the ferro-octupolar order and the reduction of the ordering temperature upon Na doping.
These findings clarify previously ambiguous experimental observations and demonstrate that charge doping in the form of small polarons offers a viable route to tuning intersite exchange interactions in spin--orbit--entangled materials, enabling the emergence of novel quantum orders.

\end{abstract}

\maketitle

Transition metal oxides with $4d$ and $5d$ magnetic ions have reshaped our understanding of strong spin-orbit (SO) coupled physics in condensed matter~\cite{Witczak2014}. These systems exhibit a wide range of phenomena, from the SO-assisted Mott insulating phase in Iridates~
\cite{Kim2008} to non trivial topology~\cite{Sinova2015} and possible Kitaev quantum spin liquids~\cite{Jackeli2009}. 
In particular, cubic double perovskites (DPs) containing $5d^1$ and $5d^2$ ions have attracted interest in the past years due to the ordering of high-rank multipoles that, although being detected experimentally, via e.g. thermodynamic measurements, 
defy precise identification due to the "blindness" of local probes~\cite{Pourovskii2025, Santini2009}. High-rank multipoles in DPs stem from SO entanglement of the spin $S$ with the effective $t_{2g}$ orbital angular momentum $l = 1$ on the 5$d$ shell. The corresponding ground state multiplet (GSM), defined by the effective total angular momentum \jeff$=S+l$, can therefore host  multipolar moments up to rank $K  = 2$\jeff~\cite{Santini2009, Kuramoto2009}. 

Different occupation of the magnetic ion  lead to distinct scenarios. 
In the case of $5d^1$, one finds  $J_{\text{eff}} = 3/2$ GSM  that can be further split in two Kramers doublets by distortions of the cubic lattice. This scenario is exemplified by Ba$_2$NaOsO$_6$, where nuclear magnetic resonance (NMR) measurements revealed two distinct phase transitions as the temperature decreases~\cite{Lu2017}. The first transition, occurring around $T_q \sim 10$ K and often referred to as “broken local point symmetry” (BLPS) phase, involves a structural distortion of the Os-O octahedra from cubic to lower symmetry, accompanied by the ordering of electric quadrupoles~\cite{Lu2017}. 
The second transition, detected at $T_N \sim$ 7 K from both magnetization measurements, muon spin resonance ($\mu$SR) and NMR, is attributed to the onset of a canted antiferromagnetic  phase with net ferromagnetic moment aligned along the [110] direction~\cite{Lu2017, Fiore_Mosca2021}.
Despite clear evidence from NMR, the exact nature of the BLPS phase remains elusive, 
as other experimental techniques have struggled to detect or characterize it~\cite{Erickson2007, Barbosa2022, Kesavan2020}, such that recent works proposed a dynamical Jahn-Teller effect as explanation for this apparent contradiction~\cite{Agrestini2024, Iwahara2023}. 

\begin{figure}[!b]
  	\begin{centering}
  	\includegraphics[width=0.85\columnwidth]{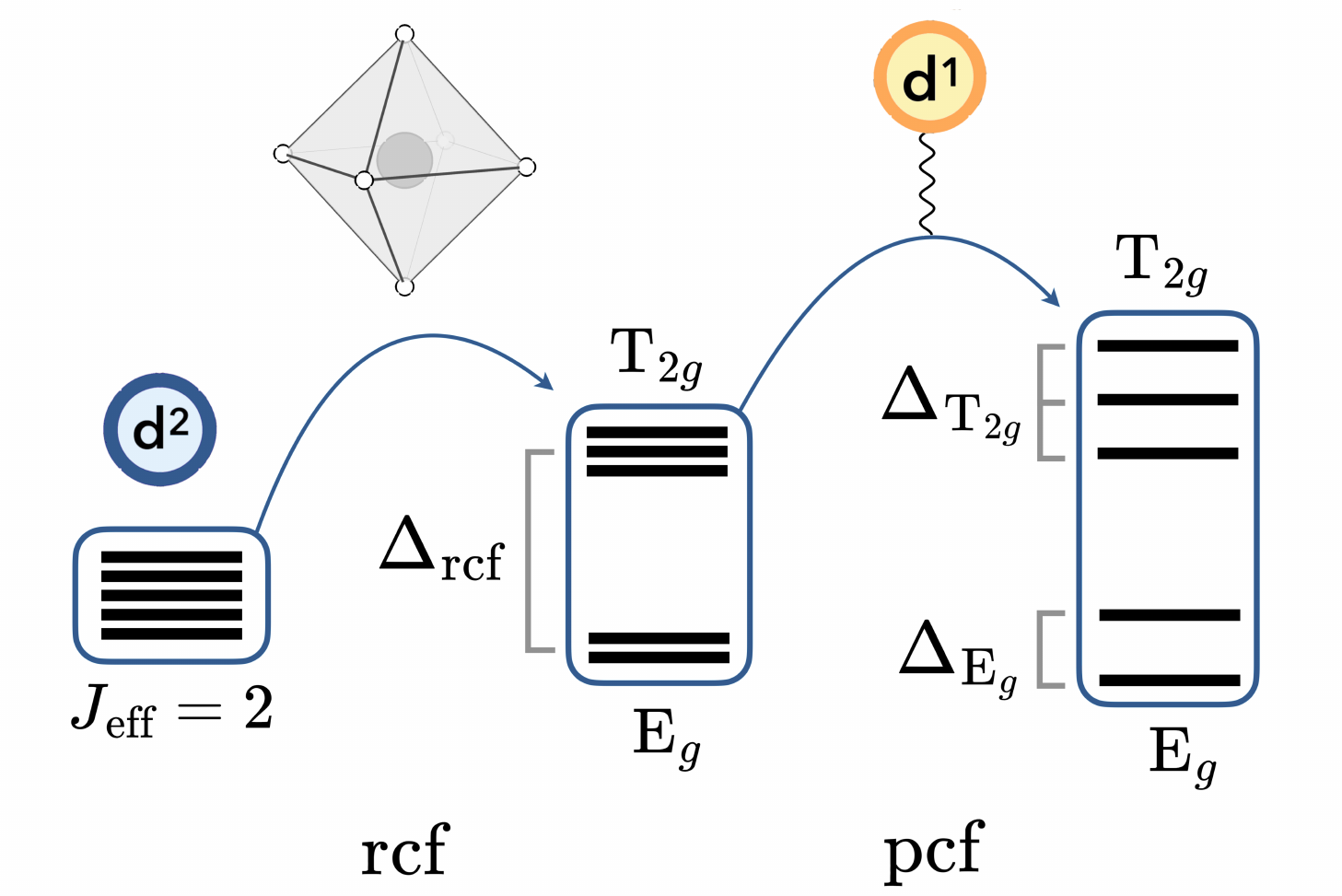} 
  		\par\end{centering}
  	\caption{Evolution of the fivefold degenerate $d^2$ GSM with \jeff= 2 (left) under  "remnant" octahedral crystal field (rcf) with $T_{2g} - E_g$ triplet-doublet splitting (middle) and polaron crystal field (pcf) acting on the non-Kramers levels causing a further splitting (right).} 
  	\label{fig:1} 
\end{figure}


Ba$_2$CaOsO$_2$, as well as isoelectronic 5d$^2$ osmates of the same family, present an equally intriguing puzzle. They exhibit a single phase transition into a state that breaks time-reversal symmetry (TRS), while lacking observable dipole magnetic moments and maintaining the cubic Fm$\bar{3}$m symmetry~\cite{Maharaj2020, Thompson2014, Marjerrison2016, Okamoto2025}.
In these systems, the $J_{\text{eff}} = 2$ GSM is split by a "remnant" octahedral CF (rcf)  in an $E_g$ non-Kramers doublet well separated in energy ($10-20$ meV) from the excited $T_{2g}$ triplet (See Figure~\ref{fig:1}). 
The $E_g$ doublet lacks a dipolar moment but can host quadrupolar and  octupolar  moments.
Theoretical and experimental works on 5$d^2$ cubic osmates  propose conflicting order parameters: either an antiferro ordering of electric quadrupoles (AF$\mathcal{Q}$)~\cite{khaliullin2021, Churchill2022}, which would break cubic symmetry while preserving TRS, or a ferro-ordering of magnetic octupoles (F$\mathcal{O}$)~\cite{Maharaj2020, Paramekanti2020, PhysRevB.101.155118, Voleti2021, Pourovskii2021, Okamoto2025}.  The latter providing a more consistent description of the experimental findings~\cite{Maharaj2020}.



In this context, 
Cong and coworkers~\cite{Cong2023} obtained valuable data on the doping-driven transition from $5d^1$ to $5d^2$ in 
\bcnoo
using magnetization measurements, $\mu$SR and NMR. 
They find the BLPS phase 
 across all NMR-accessible Na concentrations followed, upon lowering the temperature, by  a region of broken TRS, with the corresponding transition temperature $T_o$ increasing as $\delta \to 0$. 
The persistence of the BLPS phase is ascribed to \afq ordering, suggested to coexist with AFM order up to $\delta = 0.1$.  However, this interpretation conflicts with the full Ca limit ($\delta = 0$), where the $E_g$ doublet forbids dipole moments and where the mutual exclusivity of \afq and \fo phases leaves the origin of the broken TRS region unresolved. 

From a theoretical perspective, Voleti and coworkers~\cite{Voleti2023} applied a Density Functional Theory (DFT) framework to \bcnoo finding that strain fields generated by dopant Na ions in the small  $\delta\to 0$ limit
suppress the \fo transition temperature. Clusters of neighboring impurities  
are predicted to induce “parasitic” dipole moments on adjacent sites through $T_{2g}$ strain fields. 
While these results offer valuable insight into the effects of local strain on the \fo phase~\cite{Voleti2023, Voleti2021}, they do not account for the reported formation of small polarons, which underlies the coexistence of different $J_{\text{eff}}$ states~\cite{Celiberti2024}.
 


Small polarons are quasiparticles arising from the 
lattice-assisted trapping of dopant charges  \cite{Franchini2021}. In \bcnoo they promote the localization of states with distinct $5d^1$ and $5d^2$ electronic configurations, while keeping the system insulating at all $\delta$~\cite{Celiberti2024}. Though mobile at high temperatures, 
they become static at the temperatures relevant to multipolar orders. Hence, a set of static $5d^1$ polarons coupled by intersite exchanges to a matrix of $5d^2$ ions 
can give rise to novel 
ordered
phases.

In this Letter
we investigate  
the interplay of polarons 
and 
multipolar orders in 
\bcnoo 
close to the 5d$^2$ limit ($\delta \le 0.25$).
Using a fully \emph{ab initio} approach, we derive the microscopic Hamiltonian, incorporating mixed intersite exchange interactions (IEI) between $J_{\text{eff}} = 3/2$ and $J_{\text{eff}} = 2$ GSMs while accounting for CF and (hole) doping effects. Our key finding is that  the dominant IEI between $5d^1$ polarons and the $5d^2$ matrix is actually antiferro-octupolar 
(AF$\mathcal{O}$),
resulting in the occurrence of  AF$\mathcal{O}$ polaronic moments   
 within the  
broader
\fo order.
This octupolar reversal at the polaronic site is 
accompanied by small dipole moments appearing on its neighboring sites. 
The 
sign reversal stems from the  distinct physical representations in the one-electron and two-electron manifolds 
of the relevant  local operators  in the $t_{2g}$ space having opposite signs. The consequence is a sign change of  $d^2-d^1$ IEI with respect to the same coupling between ions of the same occupancy. 
Our predicted ordering temperatures show very good agreement with experimental measurements, particularly when the effects of polaron-induced structural distortions are included.


\vspace{0.2cm}
{\it Effective Hamiltonian.} The effective Hamiltonian of \bcnoo is specified by 

\begin{align}
H^{J_{\mathrm{eff}}} &= \sum_{\langle ij \rangle} \sum_{\substack{K_i K'_j \\ Q Q'}} V_{K_iK'_j}^{QQ'} (ij) O_{K_i}^{Q}(i) O_{K'_j}^{Q'}(j) \notag \\
&\quad + \sum_i H^i_{rcf} + \sum_{i \in pol_{NN}} H^i_{pcf} .
\label{eq:hamiltonian}
\end{align}
Here the first summation ($ij$) runs over the Os-Os bond, the second  over the multipolar momenta of the ranks $K_i$, $K'_j$  = 1, 2, 3 for \jeff$= 3/2$ (and 4 for \jeff$= 2$) with projections $Q$, $Q'$. The labels $K_i/K_j$ represent the total angular momentum and site-dependent ranks $K (J_i)/K (J_j)$ respectively, consequence of the site-specific GSM. For simplicity, we will omit the $i/j$ subscripts from now on. Finally, $H^i_{rcf} = V_{rcf} [O^0_4 (i) + 5 O_4^4 (i)]$ is the rcf Hamiltonian, while $H^i_{pcf} = B^0_2 O^0_2 (i) + B_2^2 O_2^2 (i)$  is the polaron crystal field Hamiltonian induced by the localized $d^1$ charge on neighboring polaronic sites (See Figure~\ref{fig:1}). 

The $J_{\mathrm{eff}}$-IEI constants $V_{KK'}^{QQ'} (ij)$ are calculated with the Force-Theorem in Hubbard-I (FT-HI) approach~\cite{Pourovskii2016}. The FT-HI derives $J_{\mathrm{eff}}$-IEI from the converged paramagnetic electronic structure obtained by DFT~\cite{Wien2k} + dynamical mean-field theory~\cite{Georges1996,Anisimov1997_1,Lichtenstein_LDApp,Aichhorn2016} in the Hubbard-I (HI) approximation~\cite{hubbard_1}.  We employ the latest version of the publicly available MagInt code, which implements the FT-HI IEI calculations for different electronic configurations and correlated environments~\cite{magint}. 
The method and computational setup are detailed in the Supplemental Material (SM)~\cite{supplmat}. 

\begin{figure}[!b]
  	\begin{centering}
  	\includegraphics[width=0.95\columnwidth]{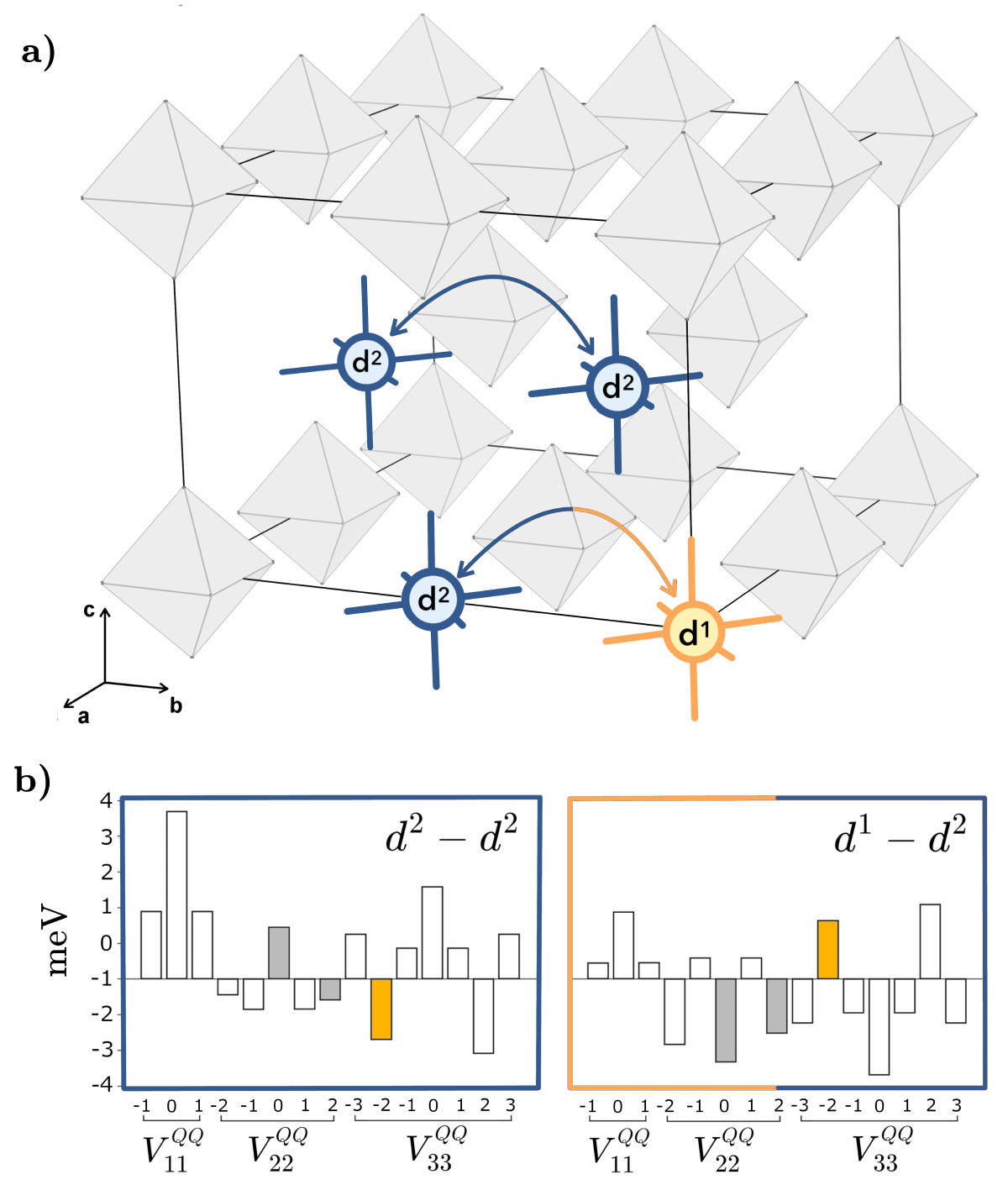} 
  		\par\end{centering}
  	\caption{a) The polaronic supercell used.  The IE paths connecting the $d^2 - d^2$ and $d^1- d^2$ are marked by blue and blue/orange arrows. b) Corresponding diagonal IEI $V_{KK}^{QQ}$. The active components within the $E_g$ GSM are indicated by filled colored bins and the octupole–octupole IE $V^{xyz}$ is highlighted in yellow. The full IEI matrices are given in the SM~\cite{supplmat}.} 
  	\label{fig:2} 
\end{figure}


\vspace{0.2cm}
{\it Polaron modeling in DFT+HI and crystal fields.}
Doping at low $d^1$ concentration is modeled  by employing the tetragonal supercell depicted in Figure~\ref{fig:2}a, containing 8 f.u.. 
To disentangle structural effects from polaronic ones, we adopt the cubic structure of the pristine $d^2$ compound of Ref.~\cite{Thompson2014} and account for hole-doping via the virtual crystal approximation.  Calculations including explicit chemical substitution for the same structure yield consistent results (see SM~\cite{supplmat}).


We find that the overall level structure arising from CF and SO effect is reproduced for both $d^1$  and $d^2$ sites. On the $d^1$ site the calculated CF and SO splittings are found to be $\approx 4.65$ eV and $\approx 0.46$ eV respectively, in good agreement with X-ray absorption near edge structure results and previous ab initio works~\cite{Kesavan2020, Fioremosca2024}. 
On the $d^2$ sites the calculated rcf splitting of the $J_{\text{eff}} = 2$ GSM  is found to be $\Delta_{rcf} \approx $ 19 meV, in very good agreement with specific heat measurements and previously reported DFT+HI calculations~\cite{Maharaj2020,Paramekanti2020, Pourovskii2021}. Both the $E_g$ and $T_{2g}$ manifolds undergo additional splittings induced by the polaron, which lowers the local $O_h$ symmetry. These pcf splittings depend on their orientation relative to the $d^1$ site: 
out-of-plane splittings reach $\Delta_{E_g} \sim 5$ meV and $\Delta_{T_{2g}} \sim 2$ meV, compared to in-plane values $\Delta_{E_g} \approx \Delta_{T_{2g}} \sim 1$ meV. 
 Additional details, wavefunctions and CF parameters are provided in the SM~\cite{supplmat}.

\vspace{0.2cm}
{\it Octupolar switching and its microscopic origin.} 
Our calculated diagonal exchange constants $V_{KK}^{QQ}$ are presented in Fig.~\ref{fig:2}b for $d^2 - d^2$ and $d^1 - d^2$ interacting bonds, as illustrated in Fig.~\ref{fig:2}a, with the full $J_{\mathrm{eff}}$-IEI matrices reported in the SM~\cite{supplmat}. As expected, mapping the $d^2 -d^2$ $J_{\mathrm{eff}}$-IEI onto the low-lying $E_g$ subspace yields a dominant octupole–octupole interaction, $V_{33}^{\bar{2}\bar{2}} \equiv V^{xyz}$, in full agreement with the pristine $d^2$ compound~\cite{Pourovskii2021}. However, a striking difference arises in the presence of polarons: while this coupling remains ferromagnetic for all $d^2 - d^2$ bonds, it reverses sign and turns antiferromagnetic for $d^1 - d^2$ bonds. This switching of the dominant multipolar channel demonstrates how the introduction of $d^1$ polarons reshapes the exchange landscape at a fundamental level, providing a microscopic origin for the destabilization of \fo order and paving the way for competing multipolar phases.

\begin{figure*}[ht]
  \begin{center}
  \includegraphics[width=0.95\textwidth, clip=true]{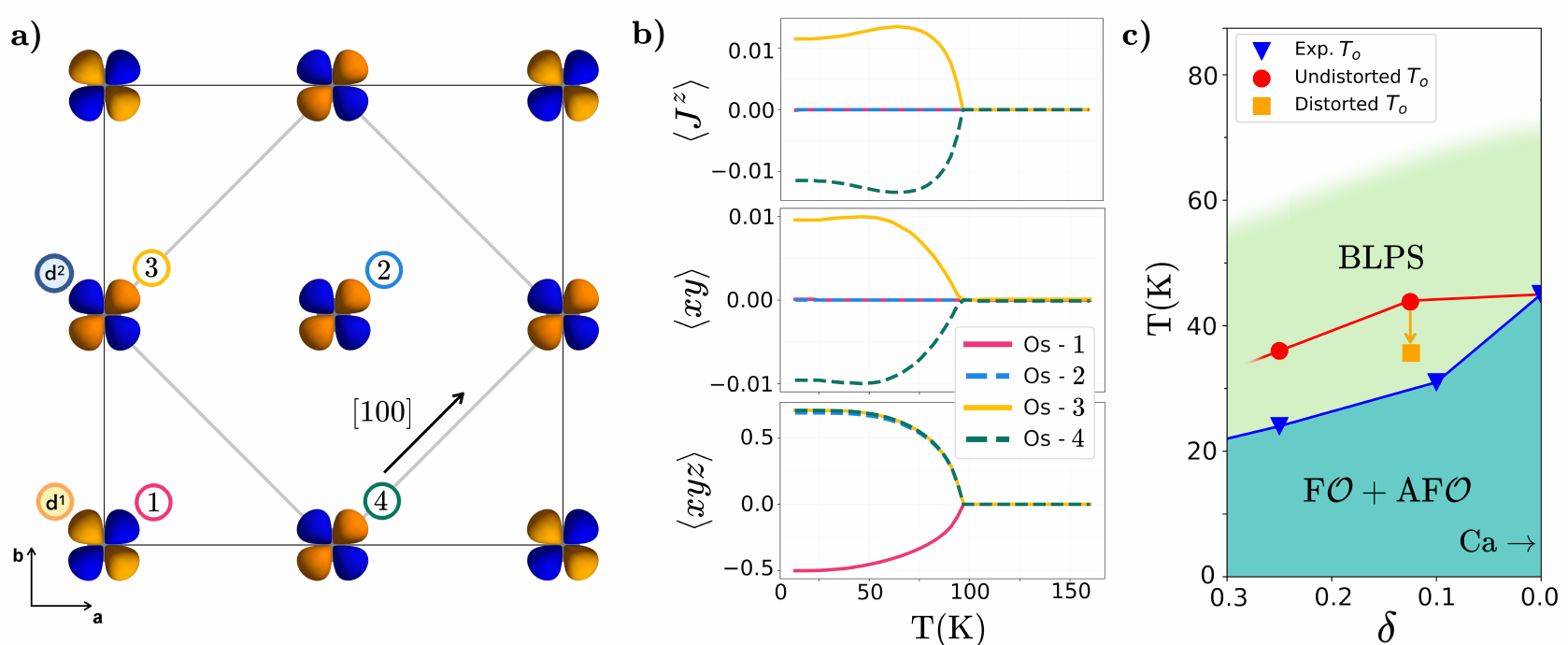}
    \end{center}
\caption{a) \fo and \afo order in \bcnoo  in the supercell used (top view). The labels 1-4 are related to the sub-figure b), where the MF values of dipole $J^z$, quadrupole $xy$ and octupole $xyz$ are shown as function of temperature.  c) Evolution of the ordering temperature $T_o$ with doping for the experimental data (blue triangles) and the undistorted cubic structure (red circles). The temperature shift at x = 0.125 arising from polaronic structural distortions is indicated by the orange square.}
\label{fig:3} 
\end{figure*}

To clarify the origin of this sign reversal, we perform a DFT+HI calculation without SO coupling in the  spin+orbital $t_{2g}$ basis.
The corresponding effective multipolar $t_{2g}$-Hamiltonian is written as:
\beq
 H^{t_{2g}} = \sum_{\langle ij \rangle} \sum_{\substack{s s' l l' \\ \alpha \alpha' \beta \beta'}}   I^{\alpha \alpha', \beta \beta'}_{s s', l l'} (ij) \tau_{s, l}^{\alpha, \beta} (i) \tau_{s', l'}^{\alpha', \beta'} (j) \ ,
\label{eq:se_hamiltonian}
\eeq
where the second summation runs over the spin ($s, s'$) and orbital ($l, l'$) ranks with components  ($\alpha, \alpha'$) and ($\beta, \beta'$) respectively, $\tau_{s, l}^{\alpha, \beta} = S^{\alpha}_{s} \otimes L^{\beta}_{l}$ are normalized spin+orbital multipolar operators~\cite{Santini2009}. 
The $t_{2g}$-shell-mediated IEI couplings ($t_{2g}$-IEI) $I^{\alpha \alpha', \beta \beta'}_{s s', l l'} (ij)$ were computed within the FT-HI approach.  

Although the $t_{2g}$-IEI are not straightforward to interpret, they can be linked to the $V_{KK'}^{QQ'}$ by SO-projection onto the \jeff many-electron bases. Specifically, we find that the relevant $t_{2g}$-IEI contributing to $V^{xyz}$ are the diagonal $I^{z z, (xy)(xy)}_{1 1, 2 2}$ terms, coupling dipole spin ($z$) and orbital quadrupole ($xy$), and corresponding permutations ($\alpha$=$x$, $\beta$=$yz$), ($\alpha$=$y$, $\beta$=$xz$).

Our SO-projection reproduces the correct sign of all $J_{\mathrm{eff}}$-IEI for both $d^2 - d^2$ and $d^1 - d^2$ bonds, but with reduced magnitude beyond dipole–dipole terms, likely due to neglecting of SOC in excited $d^1$ and $d^3$ states. Crucially, the sign reversal between $d^1 - d^2$ and $d^2 - d^2$ bonds already appears in the $I^{\alpha \alpha , \beta \beta}_{1 1, 2 2}$ couplings, showing that this effect is not induced by SO coupling. Specifically, it originates from the different representations of the one-electron $t_{2g}$-shell operators in the many-electron $d^1$ and $d^2$ bases:   the one-electron $t_{2g}$-shell operators with ($s=1$, $l=2$) acquire a minus sign when expressed in the $d^2$ many-electron Hilbert space. This sign switch is compensated on $d^2-d^2$ bonds but remains explicit on $d^1-d^2$ ones (see End Matter for the detailed derivation).

Lastly, to clarify the physical meaning  of the spin+orbital $\tau_{s,l}^{\alpha,\beta}$ operator, we express $\tau_{0,2}^{0,\beta}$ in terms of one-electron occupancy operators $n_{\gamma,\delta}$ of the $t_{2g}$ shell, as is standard in model Hamiltonian methods~\cite{Chen2010, Svoboda2021, Chen2011, Kee2023}. We find  $\tau_{0,2}^{0,xy} \propto n_{xz,yz} + n_{yz,xz}$ with $n_{yz,xz} = i(c^{\dagger}_{yz}c_{xz} - c^{\dagger}_{xz}c_{yz})$, highlighting the role of intra-orbital mixing in the emergence of high-rank multipolar phases.

\vspace{0.2cm}
{\it Ordered phase.}
We solve the full  effective Hamiltonian of Eq.~\ref{eq:hamiltonian}, incorporating both $d^1$ and $d^2$ sites, within the single site mean field (MF) approximation using the
“McPhase” package~\cite{Rotter2004} together with an in-house module. To simplify the treatment of mixed-valence interactions, we exclude the couplings to hexadecapole operators ($K = 4$)  present in both $d^1 - d^2$ and $d^2 - d^2$ $J_{\mathrm{eff}}$-IEI, as these contributions are expected to be negligible. 

We find two second-order phase transitions occurring at $T_o \sim 93$  K and $T \sim 4$ K. While we do not discuss the latter in detail here, as its very low temperature in MF makes it less significant, additional details are provided in the SM~\cite{supplmat}.
The first transition is noteworthy and leads to a complex order that combines a  \fo order of $xyz$ octupoles on all sites except on the polaronic one, where the ${xyz}$ octupole is \afo ordered (See Figure~\ref{fig:3}a). Additionally, the polaronic nearest neighbors in the $ab$-plane develop small AF magnetically ordered dipole moments of $\approx 0.01 \mu_B$. These are oriented along the $z$-axis and accompanied by the ordering of $t_{2g}$ electric quadrupoles with $xy$ character (see Figure~\ref{fig:3}b for the MF values). 

Thus, both $T_{2g}$ distortions induced by dopants~\cite{Voleti2021} and polaronic charges alter the \jeff = 2 CF to an extent that parasitic dipole moments emerge. 
Interestingly, despite the stronger out-of-plane $\Delta_{T_{2g}}$ splitting  compared to the in-plane one, no dipoles form on the out-of-plane sites. This is attributed to the ground-state wavefunctions, which do not couple to the first $T_{2g}$ excited state via the $z$-dipole operator (see SM~\cite{supplmat}). 

It is worth noting that within the $5d^1$ \jeff = 3/2 manifold all multipolar degrees of freedom are, in principle, active and can order.
Our results demonstrate that the rcf on the $d^2$ sites quenches all multipolar IE channels except those allowed within the $E_g$ doublet, thereby imposing an "effective" CF on the polaronic site.

\vspace{0.2cm}
{\it Ordering Temperature evolution} We investigate how hole doping affects the transition temperature in \bcnoo. The outcome is not obvious apriori: although the $d^1 - d^2$ $J_{\mathrm{eff}}$-IEI interactions can be up to $\sim 80$ \% weaker than in the pristine $d^2$ case~\cite{Pourovskii2021}, the $V^{xyz}$ coupling remains comparable (see the SM~\cite{supplmat}). To track the evolution of the ordering  temperature, we compute the $J_{\mathrm{eff}}$-IEI at higher Na concentration $\delta=0.25$.

The model averages two polaronic geometries, with the extra 
$d^1$ charge residing either in-plane (next-nearest neighbor) or out-of-plane (nearest neighbor).
The corresponding MF solutions 
contain
two \afo ordered polarons in both cases, confirming the robustness of octupolar order against charge doping at these low Na concentrations.
After rescaling the MF values to match the experimental $T_o$ at $\delta = 0$, we find that the undistorted results already reproduce the overall experimental trend (Fig.~\ref{fig:3}c, red circles). The reduction of $T_o$ upon doping arises from a subtle interplay of several factors: polaron-induced changes in the $E_g - T_{2g}$ rcf splittings, modifications of the pcf at higher $\delta$, slightly weakened $xyz$ IE and the different multiplicity of the magnetic ion’s GSMs within the supercell (see End Matter).

Structural distortions are also expected to play an essential role, since polaron localization necessarily involves local lattice relaxation. To examine this effect, we performed a full relaxation of the structure for $\delta = 0.125$ 
using the Vienna Ab Initio Simulation Package (VASP)~\cite{PhysRevB.54.11169} within the constrained DFT+U method. The on-site density matrices were initialized with the combined \fo+\afo\ configuration, as obtained at DFT+HI level.
The approach, detailed in Ref.~\cite{FioreMosca2022} and summarized in the SM, showed to accurately capture high-rank multipolar phases in DFT. From the optimized distorted structure we computed both $J_{\mathrm{eff}}$-IEI and ground state properties in FT-HI. 

The MF calculation reproduce the previously identified \afo polaron embedded in the \fo matrix, albeit with stronger parasitic dipole moments, now distributed across all osmium sites (see SM~\cite{supplmat}). Moreover, we find that including polaron-induced structural effects improves the agreement with experiment by a large extent (Fig.\ref{fig:3}c, orange square). It is  therefore reasonable to expect that chemical substitution and a more complex treatment of Na dopant configurations, as examined in Ref~\cite{Voleti2023}, could lead to an even closer quantitative agreement with the experimental data.

\vspace{0.2cm}
{\it Conclusions.} 
To bridge the distinct physical regimes of $5d^1$ and $5d^2$ double perovskites and connect their different J-effective multipolar phases, we derived and solved from first principles the effective Hamiltonian of
\bcnoo, explicitly accounting for polaronic charges at low Na-doping levels through material-specific DFT calculations.
Our analysis demonstrates a polaron-induced reversal of the dominant octupolar exchange, originating from different representation of $t_{2g}$-shell spin+orbital operators in one-electron and %
two-electron bases. This reversal of the octupolar IEI provides direct evidence for a coupling between polaronic degrees of freedom and magnetic interactions, an effect not previously reported. Beyond explaining the experimentally observed suppression of the ordering  temperature with doping, this result opens a pathway to designing and controlling novel quantum orders in spin–orbit systems through charge doping.

\hspace{0.1cm}

\begin{acknowledgements}
Support by the the Austrian Science Fund (FWF) grant J4698 is gratefully  acknowledged.  
D.~F.~M. and L.~C. thank the computational facilities of the Austrian Scientific Computing (ASC). L.~V.~P. acknowledges the support by CNRS through the Tremplin@Physique 2025 program and  by the  CPHT computer team.
The authors thank the Erwin Schrödinger Institute (ESI) for hosting the ESI-PsiK workshop
‘Spin–orbit entangled quantum magnetism’ and all participants for the many enlightening discussions.
\end{acknowledgements}

\bibliography{bibliography}

\clearpage

\section{End Matter}

\subsection{From $t_{2g}$- to $J_{\mathrm{eff}}$- Hamiltonian} 

The $t_{2g}$-Hamiltonian of Eq.\ref{eq:se_hamiltonian} is defined in the spin+orbital many-electron basis states $|\tilde{\Gamma} (i) \rangle$, which, for the two-electron case, belong to the Hund's rule GS multiplet with the degeneracy $(2l+1)(2S+1)=9$ for $S=1$ and $l=1$. 
The corresponding intersite exchange interactions ($t_{2g}$-IEI) can be downfolded onto the \jeff many-electron basis $|\Gamma (i) \rangle$. Here $i$ is the magnetic site index (1 for $d^1$ and 2 for $d^2$ for simplicity). To this end, we introduce the projector operator $P_{LS} (i) = |\Gamma (i)  \rangle \langle \tilde{\Gamma} (i)|$, such that any many-electron $t_{2g}$-shell operator $\tau_{s, l}^{\alpha, \beta}$  can be $J_{\mathrm{eff}}$-projected as 
\beq
   \tilde{O}^{\alpha, \beta}_{s,l} (i)  = P_{LS} (i)\  \tau_{s, l}^{\alpha, \beta} (i) \ P^\dagger_{LS}(i).
\eeq
The multipole many-electron operators $O_K^Q$ are expanded in terms of the projected operators through coefficients   $c_{s\alpha, l\beta}^{KQ} (i) = Tr[\tilde{O}^{\alpha, \beta}_{s,l}(i) O_K^Q]$. Specifically, for the $(K,Q)=(3,-2)$ component, corresponding to the $xyz$ octupole in Cartesian coordinates, this expansion yields the values listed in Tab.~\ref{tab:expansion} for both $d^1$ and $d^2$ configurations.

\begin{figure}[!t]
  	\begin{centering}
  	\includegraphics[width=0.95\columnwidth]{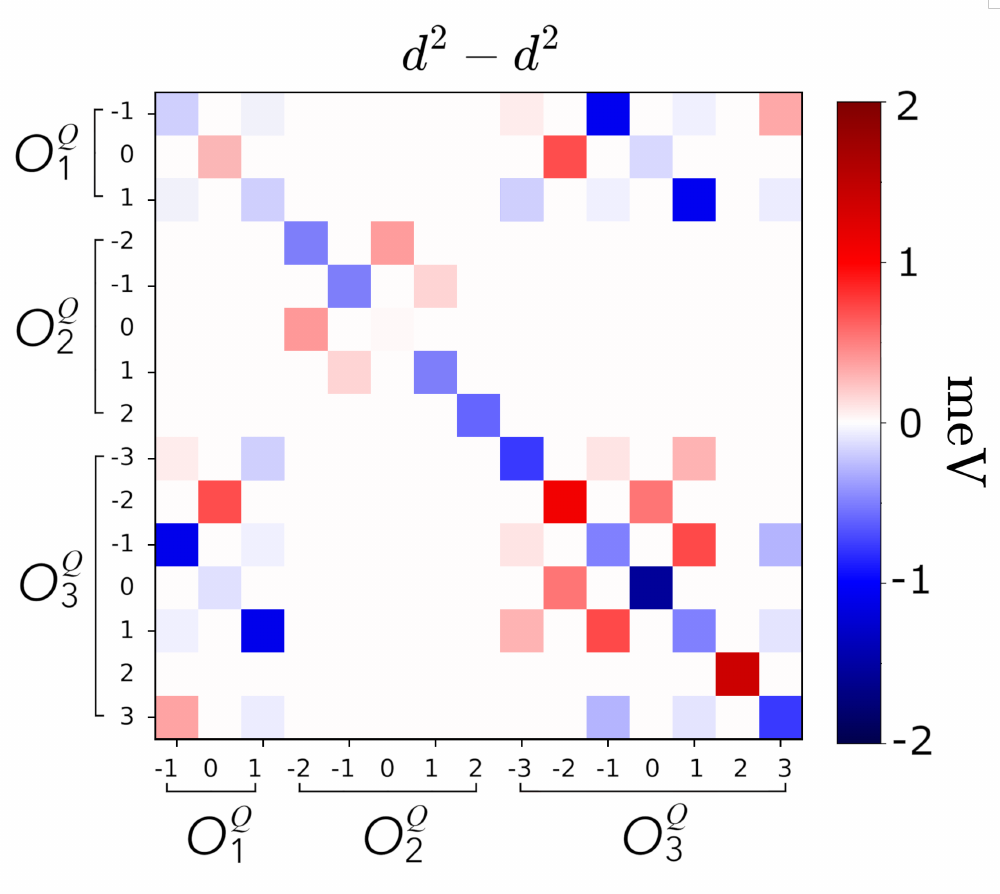} 
  		\par\end{centering}
  	\caption{Difference between the SO-projected $t_{2g}$-IEI and the IEI with DFT+SO for the $d^2-d^2$ interaction bond $\mathbf{R}_{ij} = [0,0.5,0]$ in the supercell reference frame.} 
  	\label{fig:endmatter1} 
\end{figure}

The $J_{\mathrm{eff}}$-IEI are then obtained by contracting these coefficients with the original interaction matrix elements:
\beq 
V^{QQ'}_{KK'} (ij) = \sum_{s,l}\sum_{\alpha, \beta}  I^{\alpha \alpha', \beta \beta'}_{s s', l l'} (ij) c_{s\alpha, l\beta}^{KQ} (i) c_{s\alpha, l\beta}^{K'Q'} (j). 
\eeq
The difference between the resulting SO-projected $t_{2g}$-IEI matrix and the initial $J_{\mathrm{eff}}$-IEI is shown in Fig.~\ref{fig:endmatter1}. The SO-projected IEI reproduce the overall structure (See SM~\cite{supplmat}), with excellent agreement in the dipole–dipole channels, while higher-order multipolar terms are systematically underestimated. For instance, the octupole–octupole coupling $V^{xyz}$ is $\sim -0.6$~meV for the $d^2$–$d^2$ bond, compared to the expected $-1.6$~meV. This discrepancy likely originates from neglecting SO coupling in the excited $d^1$ and $d^3$ configurations in the superexchange processes, though further investigation is required.
\vspace{0.5cm}
\begin{table}[h]
\centering
{\renewcommand{\arraystretch}{1.2}
\begin{tabular*}{0.7\columnwidth}{@{\extracolsep{\fill}}llcc}
\hline\hline
\multicolumn{2}{c}{SE operator} &  \multicolumn{2}{c}{$c_{s\alpha, l\beta}^{KQ}$} \\
\hline
$s, \alpha$ & $l, \beta$ & $d^1$  & $d^2$ \\
\hline
$1, x \ \ \ $ & $2, yz\ \ \ $ & -0.5757 & -0.3826 \\
$1, y \ \ \ $ & $2, xz\ \ \ $ & -0.5757 & -0.3826 \\
$1, z \ \ \ $ & $2, xy\ \ \ $ &  0.5757 &  0.3826 \\
\hline\hline
\end{tabular*}}
\caption{Expansion coefficients for the octupolar operator $O_3^{-2}$ in the spin+orbital basis.}
\label{tab:expansion}
\end{table}


\subsection{Origin of sign switch}
As mentioned in the main text, we observe a reversal of the IEI already without SO coupling and within the $t_{2g}$-shell treatment. This is shown in Tab.~\ref{tab:matrices}, where the $t_{2g}$-IEI are written, for the $t_{2g}$-shell operators leading to the $V^{xyz}$ coupling.

To understand the origin of the sign switch, we compute the matrix elements of the $d^1$ one-electron $t_{2g}$-shell operator  $\tau_{s, l}^{\alpha, \beta} (1)$ in the many-electron $d^2$ spin+orbital basis via 
\beq
[T_{s, l}^{\alpha, \beta} (2)]_{MM'} = \langle \Tilde{\Gamma}(2)M | \tau_{s, l}^{\alpha, \beta} (1) | \Tilde{\Gamma}(2) M'\rangle .
\eeq
We then express the resulting operator in terms of $d^2$ many-electron operators $\tau_{s', l'}^{\alpha', \beta'} (2)$ by fitting its components through $t_{ss', ll'}^{\alpha \alpha', \beta\beta'}  = Tr[\tau_{s', l'}^{\alpha', \beta'} (2) T_{s, l}^{\alpha, \beta} (2)] $. In the expansion, we find the correct matrices of 2-electrons operators ($s=s'$, $l=l'$, $\alpha = \alpha'$, $\beta = \beta'$), but with different signs for different components: 

i) $t_{ss, ll}^{\alpha \alpha, \beta\beta} = -1$ for $(s,l) = (0, 2)$, i.e. for pure orbital-quadrupole terms.

ii) $t_{ss', ll'}^{\alpha \alpha, \beta\beta} = 1$ for both $(s, l) = (0, 1)$ dipole-orbital terms and $(s, l) = (1, 0)$ spin-dipole terms. The sign reversal thus arises in cases where multiplicative contributions remain uncompensated; for instance, when a spin-dipole coupled to an orbital-quadrupole on a $d^2$ site interacts with the corresponding spin-dipole orbital-quadrupole term on a $d^1$ site.

\begin{table}[h]
\centering
{\renewcommand{\arraystretch}{1.2}
\begin{tabular*}{\columnwidth}{@{\extracolsep{\fill}} c|ccc |ccc}
\hline\hline
 \multicolumn{1}{c}{} &  \multicolumn{3}{c}{\textbf{$d^2 - d^2$}} & 
\multicolumn{3}{c}{\textbf{$d^1 - d^2$}} \\
\hline 
  & $x;yz$ & $y;xz$ & $z;xy$ &  $x;yz$ & $y;xz$ & $z;xy$ \\
\hline
$x;yz$ & -2.62 &       &       &  2.97 &       &       \\
$y;xz$ &       & -2.62 &       &        &  2.97 &       \\
$z;xy$ &       &       &  1.23 &       &       & -1.39 \\
\hline\hline
\end{tabular*}}
\caption{$t_{2g}$-IEI couplings for $\tau_{s,l}$ with $(s,l)=(1,2)$, i.e. spin-dipole and orbital-quadrupole operators. The corresponding $(\alpha,\beta)$ components are listed. These couplings project onto the spin–orbit $V^{xyz}$ octupole–octupole intersite exchange. Values are reported for $\mathbf{R}_{ij} = [0,0.5,0]$ in the supercell reference frame of Fig.~\ref{fig:2}a.}
\label{tab:matrices}
\end{table}

\subsection{Mean Field analysis of $T_o$}

A detailed analysis of the evolution of $T_o$ with Na concentration is necessary. As discussed in the main text, the $V^{xyz}$ octupolar interaction changes only weakly between $x = 0$ and $x = 0.125$. Yet the ordering temperature decreases noticeably, especially at higher doping levels ($x = 0.25$) in the undistorted cubic structure. Interestingly, at these Na concentrations, despite the sign change in the IEI, the system remains non-frustrated: reversing both the interactions and the  order parameters leaves the total energy unchanged.

\begin{figure}[!t]
  	\begin{centering}
  	\includegraphics[width=0.95\columnwidth]{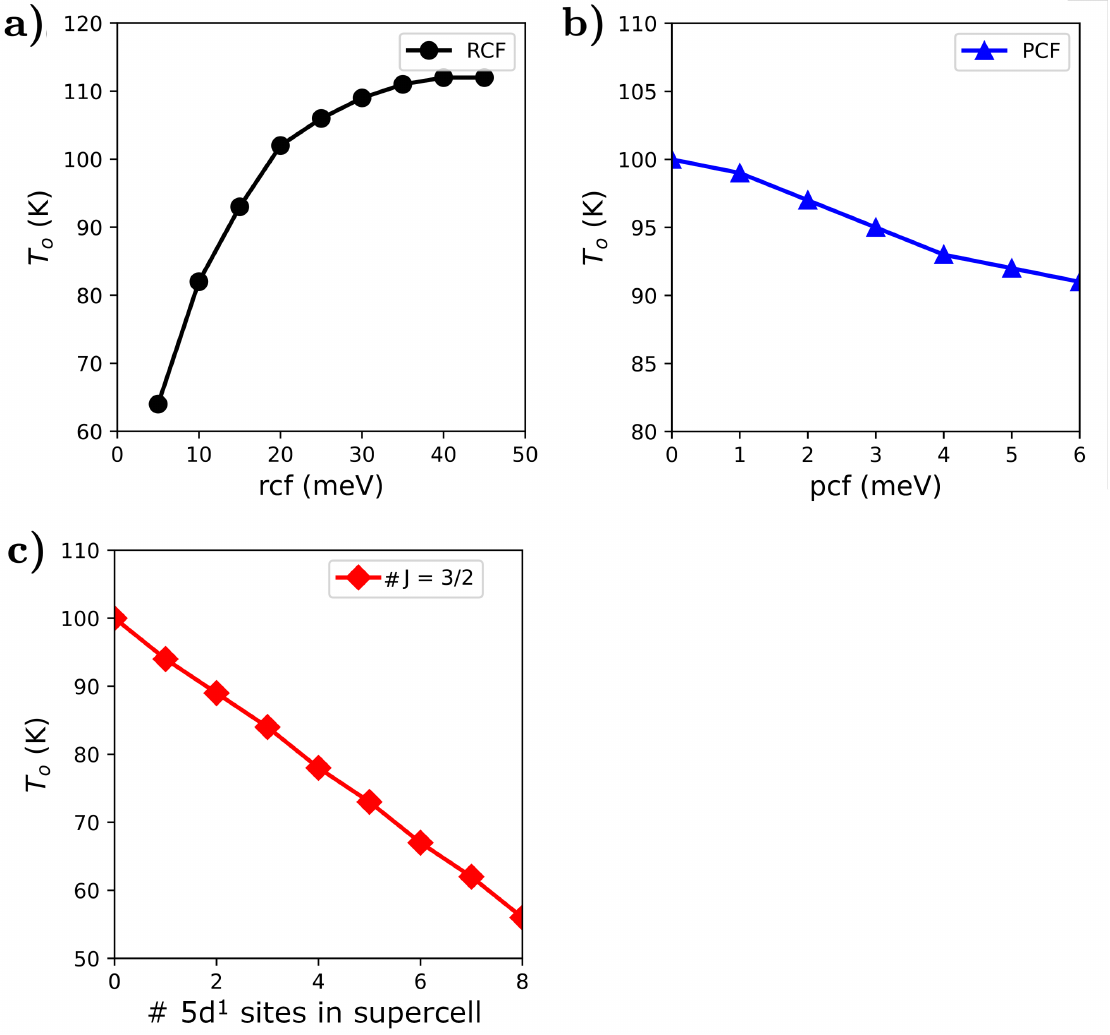} 
  		\par\end{centering}
  	\caption{Mean-field ordering temperature $T_o$ as a function of (a) remnant crystal field (rcf) within the \jeff = 2 ground-state multiplet (GSM), and (b) polaron crystal field (pcf) for the IEI computed in the full $d^2$ BCOO compound. (c) $T_o$ as a function of the number of hole-polarons (\jeff = 3/2 GSMs) in the supercell with only the $V^{xyz}$ octupole-octupole IEI active.} 
  	\label{fig:endmatter2} 
\end{figure}

To isolate the microscopic origin of the lowering of $T_o$ upon Na doping, we performed a \emph{gedankenexperiment} in which we fixed IEI for the \fo ordered pristine case ($\delta = 0$), while varying the rcf and pcf splittings incrementally (See Figure~\ref{fig:endmatter2}a,b). The results show a strong dependence of $T_o$ on the rcf, saturating at $\sim  45$ meV to $\approx 112$ K and decreasing to roughly half the value when the rcf is reduced to $\sim 6$ meV. Any lower rcf makes the $T_{2g}$ levels magnetically active and promotes dipolar order. In contrast, the pcf exerts only a weak, nearly linear influence: increasing it up to 6 meV lowers $T_o$ by only a few percent. Higher values were not explored, as stronger pcf terms would induce significant $T_{2g}$ mixing in the $E_g$ doublet and alter the qualitative physics.

While variations in the rcf and pcf account for part of the observed suppression of $T_o$, they do not capture the full picture. An additional factor is the difference in the total angular momentum GSMs between the $d^2$ (\jeff= 2) and $d^1$ (\jeff = 3/2) configurations. Indeed, as seen in Fig.~\ref{fig:endmatter2}a, even for rcf values of $\sim 20$ meV, the \jeff = 2 GSM does not yet fully project onto a pseudospin-1/2 subspace, with clear implications in the ordering behavior.

To assess the effect of GSM mismatch, we performed a second \emph{gedankenexperiment} in which all intersite exchange interactions except the octupole-octupole $V^{xyz}$  were set to zero. The doping concentration was then varied by progressively substituting $d^2$ sites with $d^1$ ones, effectively replacing the \jeff = 2 GSM (with its $\sim 20$ meV rcf) by the \jeff = 3/2 GSM (Fig.~\ref{fig:endmatter2}c). This procedure preserves the stability of the \fo solution across all doping levels and isolates the role of the GSMs. The resulting $T_o$ decreases linearly with increasing $d^1$ concentration, again in agreement with the trend observed both in our calculations and in experimental observations.

Additional mechanisms, such as static or dynamic Jahn–Teller effects on the $d^1$ polarons, or $d^1 – d^1$ quadrupole–quadrupole dimerization and dopant-induced strains may also affect the ordering temperature, but their investigation is beyond the scope of the present work.

\clearpage

\end{document}


\title{Supplementary material for 
'Polaron-driven switching of octupolar order in doped 5d$^2$ double perovskite'}

\author{Dario Fiore Mosca}
\address{University of Vienna, Faculty of Physics, Center for Computational Materials Science, Kolingasse 14-16, 1090, Vienna, Austria}
\address{CPHT, CNRS, \'Ecole polytechnique, Institut Polytechnique de Paris, 91120 Palaiseau, France}
\address{Coll\`ege de France, Université PSL, 11 place Marcelin Berthelot, 75005 Paris, France}

\author{Lorenzo Celiberti}
\address{University of Vienna, Faculty of Physics, Center for Computational Materials Science, Kolingasse 14-16, 1090, Vienna, Austria}

\author{Leonid V. Pourovskii}
\address{CPHT, CNRS, \'Ecole polytechnique, Institut Polytechnique de Paris, 91120 Palaiseau, France}
\address{Coll\`ege de France, Université PSL, 11 place Marcelin Berthelot, 75005 Paris, France}

\author{Cesare Franchini}
\address{University of Vienna, Faculty of Physics, Center for Computational Materials Science, Kolingasse 14-16, 1090, Vienna, Austria}
\address{Department of Physics and Astronomy "Augusto Righi", Alma Mater Studiorum - Universit\`a di Bologna, Bologna, 40127 Italy}

\setlength{\parindent}{0pt}

\date{\today}	
\maketitle

\section{First principles methods}

In the following, we describe our first principles calculations. In A. we detail the electronic structure calculation of paramagnetic \bcnoo with both virtual approximation and chemical substitution, in B. the calculation of Intersite Exchange Interactions (IEI) for both the spin-orbit \jeff ground state multiplets (GSMs) and ($s, l$) spin+orbital $t_{2g}$-shell. In C. we detail the modeling of octupolar polaron in constrained density functional theory (DFT) + U.

\subsection{Correlated electronic structure calculations}\label{ssec:dft+hi}

We calculate the paramagnetic electronic structure of \bcnoo  using the charge self-consistent DFT~\cite{Wien2k} + dynamical mean-field theory (DMFT)~\cite{Georges1996,Anisimov1997_1,Lichtenstein_LDApp,Aichhorn2016} within the quasi-atomic Hubbard-I (HI) approximation~\cite{hubbard_1} based on the full-potential LAPW code Wien2k~\cite{Wien2k}, incorporating spin-orbit (SO) coupling via the standard variational treatment.

To model doping we utilize an 8 f.u. supercell obtained from the cubic Ba$_2$CaOsO$_6$ structure, with lattice parameters $\sqrt{2}a \times \sqrt{2}a \times a$ and $a = 8.346$ \AA~\cite{Thompson2014}. Doping concentration $\delta = 0.125$ is obtained using the virtual crystal approximation, achievable by reducing the total electron count by one. Additional calculations incorporating  explicit chemical doping, as presented in Sec. IV, have been done by explicitly substituting the Ca ion at position [0, 0, 0.5] by one Na. 

We employ the local density approximation (LDA) for the DFT exchange-correlation potential, a 30 $\vk$-point mesh across the full Brillouin zone and a basis cutoff of $R_{mt}K_{max} = 7$.  For the correlated space projection, we consider the Wannier orbitals representing the full Os $d$ states constructed from the manifold of Kohn-Sham bands within the energy window $[-1.36:5.44]$~eV relative to the Fermi level. 

The polaronic $5d^1$ localization in DFT+HI is enforced via the double-counting correction in the fully localized limit by explicitly setting the $d^1$ electron count on the polaronic site, while nominal $d^2$ are maintained for all other Os sites. For the DFT+HI calculation, we employ the $d$-shell parameters $F^0 = U = 3.2$ eV and $J_{H} = 0.5$ eV on all sites, consistent with previous studies on $d^1$ and $d^2$ DPs~\cite{Fiore_Mosca2021,Pourovskii2021}.

\subsection{Calculation \jeff-IEI  and $t_{2g}$-IEI}

The multipolar IEI of both  Eq. 1 (Main) and Eq. 2 (Main) are determined using the Force-Theorem in HI (FT-HI) method described in Ref~\cite{Pourovskii2016}. This approach involves inducing small symmetry-breaking fluctuations in the density matrix of the GSMs occurring simultaneously at two neighboring magnetic (Os) sites, $i$ and $j$. The interaction terms $V^{KK’}_{QQ'}(ij)/I_{ss', ll'}^{\alpha \alpha', \beta \beta'}$ are then extracted by analyzing how the DFT+DMFT grand potential responds to these two-site fluctuations. Detailed explanations can be found in the Appendix of Ref.~\cite{Pourovskii2023} and the Supplementary Material of Ref.~\cite{Fiore_Mosca2021}, with the full derivation provided in Ref.~\cite{Pourovskii2016}.
The reference frame for the calculated IEI is aligned with the local octahedral axes shown in Figure 2b (Main), which are rotated $45^\circ$ relative to the cubic crystallographic [100] axis. When necessary, like for physical magnetic moments in the global crystallographic frame, a further $45^\circ$ rotation is applied.

\subsection{Modeling of octupolar polaron in DFT + U}

The polaronic localization is obtained at DFT+U level by adopting the constrained initialization of the on-site density matrix (ODM) as introduced in Ref.~\cite{FioreMosca2022}. This method constructs an ODM for each site based on the order parameters of the \fo and \afo polaronic multipolar phases. In the \jeff space, the ODM is expressed as:
\beq
\rho^{mm'}_{HI} = Tr[\rho^{mm'}_{MM'} (J_{\text{eff}}) \hat{\rho_{\alpha}} (J_{\text{eff}})]
\eeq
where $\rho^{mm'}_{MM'} (J_{\text{eff}})$ is the matrix element of the ODM operator in the GSM basis and $\hat{\rho_{\alpha}} (J_{\text{eff}})$ is the many-electron density matrix derived from the GSM for a given phase. For instance, to initialize the octupolar polaron configuration, we define  $\hat{\rho_{\alpha}} = \hat{O}_{xyz} c_{xyz}$ for all $d^2$ sites and $\hat{\rho_{\alpha}} = - \hat{O}_{xyz} c_{xyz}$ for the polaronic $d^1$ site, where $c_{xyz} \equiv \langle \hat{O}_{xyz} \rangle $ is the expectation value  of the $xyz$ octupole at site $\alpha$.

The many electron DM $\hat{\rho_{\alpha}} (J_{\text{eff}})$ contains, by construction,  the nominal number of correlated electrons included in the effective low-energy Hamiltonian. Before the DFT+U self-consistent cycle, this number is corrected to account for the strong hybridization effects that enhance $5d$ orbital occupations in double perovskites~\cite{Agrestini2024, FioreMosca2022}. We find the total electron number in our DFT+U ODMs to be $\approx$ 5.95 and $\approx$ 5.93 for $d^2$ and $d^1$ sites respectively. Further details of this procedure can be found in Ref.~\cite{FioreMosca2022}. The initialized ODMs can be found in Sec. V.

The DFT+U calculations were performed using the Vienna Ab Initio Simulation Package (VASP)~\cite{PhysRevB.54.11169,PhysRevB.47.558}, employing the Perdew-Burke-Ernzerhof (PBE) generalized gradient approximation. The on-site Coulomb repulsion at the Os $d$ orbitals was applied using the rotationally-invariant Lichtenstein formulation of DFT+U~\cite{PhysRevB.52.R5467}. We used the on-site Coulomb $U = 3.2$ eV and $J_H = 0.5$ eV, consistent with prior studies~\cite{FioreMosca2022, Pourovskii2021}.  The reciprocal space was sampled with a k-point spaced mesh of 0.1 \AA$^{-1}$, the plane wave energy cutoff of 550 eV was used together with an energy con convergence of $10^{-4}$ eV. The structural relaxation was performed using the Blocked-Davidson algorithm,
allowing for the optimization of cell volume, cell shape, and atomic positions with a convergence accuracy of 10$^{-2}$ \AA/eV on the atomic forces.

\clearpage
\section{Stability of IEI and Methodological Considerations}

\begin{figure}[!h]
  	\begin{centering}
  	\includegraphics[width=0.7\columnwidth]{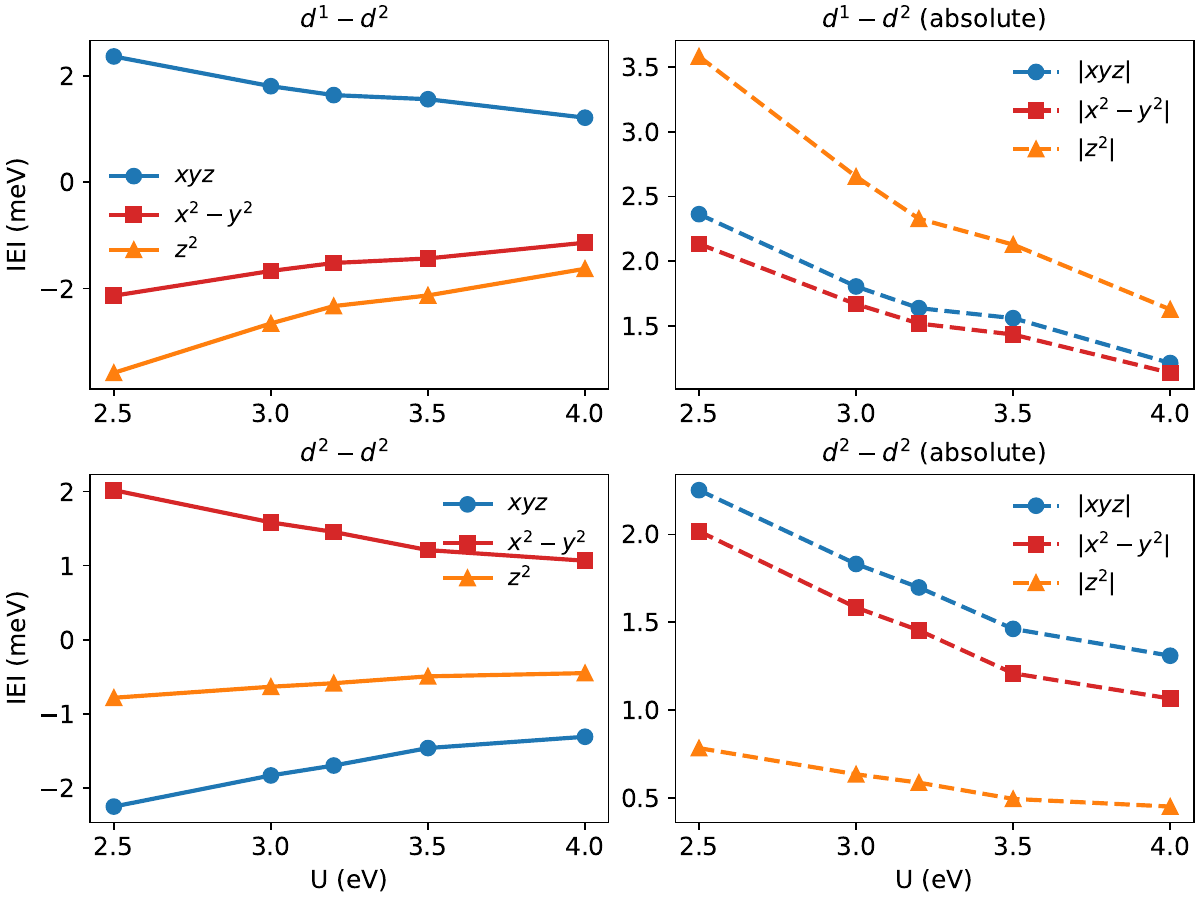} 
  		\par\end{centering}
  	\caption{$V^{xyz}$ ($xyz$), $V^{x^2-y^2}$ ($x^2-y^2$) and  $V^{z^2}$ ($z^2$) IEI as a function of U for the $\mathbf{R} = [0, 0.5, 0]$ bond in the supercell reference frame for (a) $d^1 - d^2$  and (c) $d^2 - d^2$ interactions, and (b, d) the respective absolute values.} 
  	\label{fig:suppl_u} 
\end{figure}

We have investigated the dependence of the IEI on the Hubbard interaction U by performing a series of calculations in which U is varied from 2.5 to 4.0 eV in steps of 0.5 eV. Figure~\ref{fig:suppl_u} shows the resulting evolution of the octupole–octupole ($V^{xyz}$) and quadrupole–quadrupole ($V^{x^2-y^2}$, $V^{z^2}$) IEI for the $\mathbf{R}=[0,0.5,0]$ bond in the supercell reference frame, for both $d^1 - d^2$ and $d^2 - d^2$ pairs.

Across the entire range of U values considered, the qualitative behavior of the exchange interactions remains unchanged. In particular, the octupole–octupole interaction $V^{xyz}$ on the $d^2$ sites remains the dominant contribution  and its sign and relative strength are preserved. Consequently, the switching of the octupolar order induced by the presence of polarons, the main result of this work, is robust with respect to variations of U. Changes in U within the physically relevant range do not modify the underlying multipolar ordering tendencies on either the $d^2$ or the $d^1$ sublattices.

Recently, symmetry-breaking approaches such as special quasirandom structure (SQS) models have been proposed to capture insulating behavior~\cite{Zunger2020} and have been applied, among other systems, to heavy transition-metal oxides~\cite{Kim2016,Liu2016}.

Osmium double perovskites, however, present a particular challenge for this approach. Their magnetically ordered phase is characterized by multipolar order, in which conventional magnetic dipole moments are absent, complicating the definition of a physically meaningful paramagnetic reference state. Moreover, studies of 5$d^1$ osmium double perovskites have shown that simulations based on randomly disordered local spin orientations require an explicit on-site Hubbard-$U$ correction to open an insulating gap~\cite{Fioremosca2024,Tehrani2021}. This indicates that these systems cannot be regarded as genuine local-moment paramagnetic insulators. Instead, their insulating behavior appears to emerge from a nontrivial interplay of electronic correlations, spin–orbit coupling, and lattice-symmetry breaking; a scenario that warrants further investigation.

\clearpage
\section{Doping concentration $\delta = 0.125$ with virtual crystal approximation}

In the following, we present DFT+HI, FT-HI, and mean-field (MF) results for \bcnoo within the virtual crystal approximation at Na doping concentration $\delta = 0.125$. Section A describes the structural details, Section B reports the DFT+HI Density of States (DOS), Section C discusses the $J_{\mathrm{eff}}$-IEI, Section D analyzes the crystal-field splittings and ground-state multiplet wavefunctions, and Section E presents the ordered phase and the MF order parameters as a function of temperature.  Lastly, Section F presents the spin+orbital $t_{2g}$-IEI.

\subsection{Structure}

\begin{figure}[!h]
    \centering
    \includegraphics[width=0.45\textwidth]{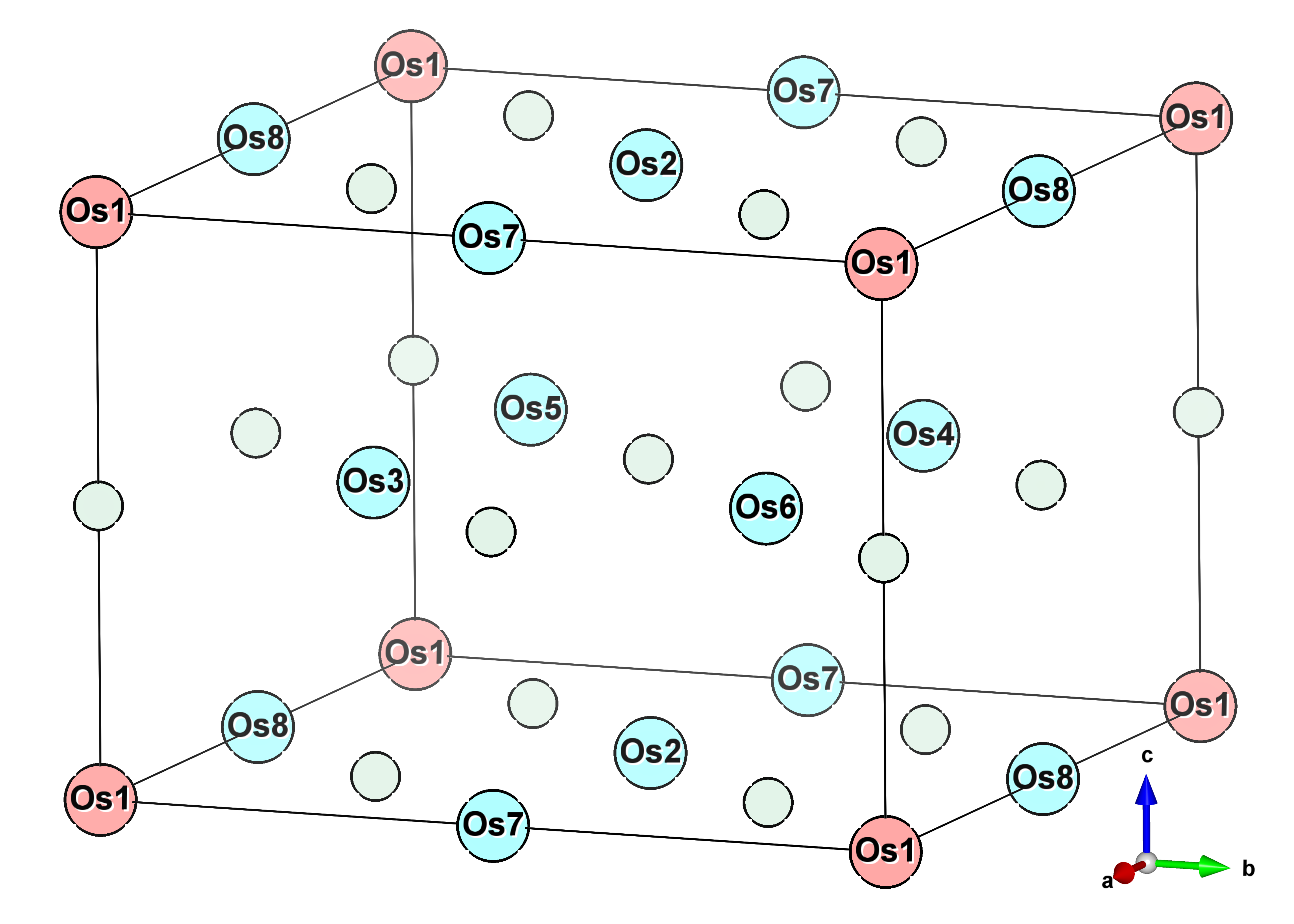}
    \caption{Supercell of \bcnoo with labels for the osmium ions as indexed in the next Sections. The polaron is localized on the Os1 (light red).}
    \label{fig:struct1}
\end{figure}

\subsection{Polaronic density of States in DFT, DFT+HI}

\begin{figure}[!h]
    \centering
    \includegraphics[width=0.45\textwidth]{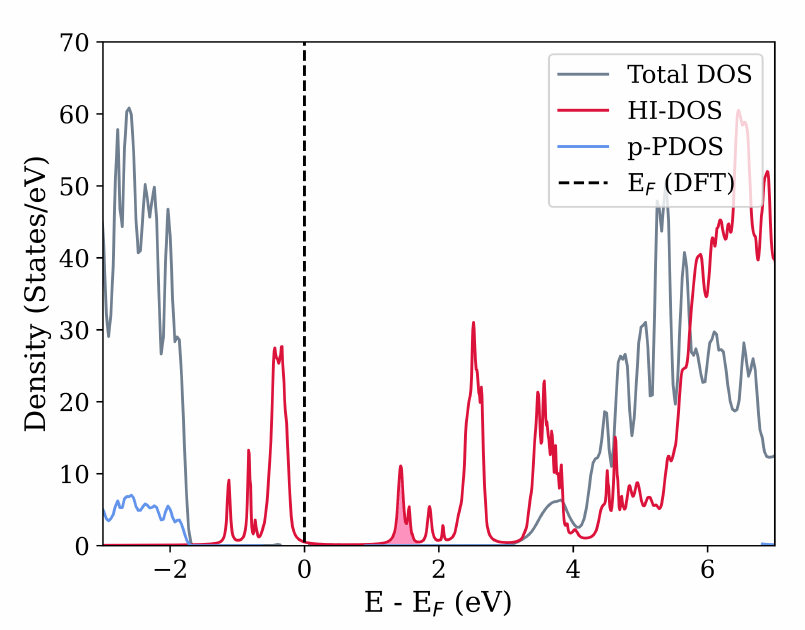}
    \caption{Combined the DFT and DFT+HI DOS. The hole-polaron peak is highlighted in light-red.}
    \label{fig:dos1}
\end{figure}

\subsection{Intersite Exchange Interactions in J$_{\mathrm{eff}}$ basis}

\begin{figure}[!h]
    \centering
    \includegraphics[width=0.9\textwidth]{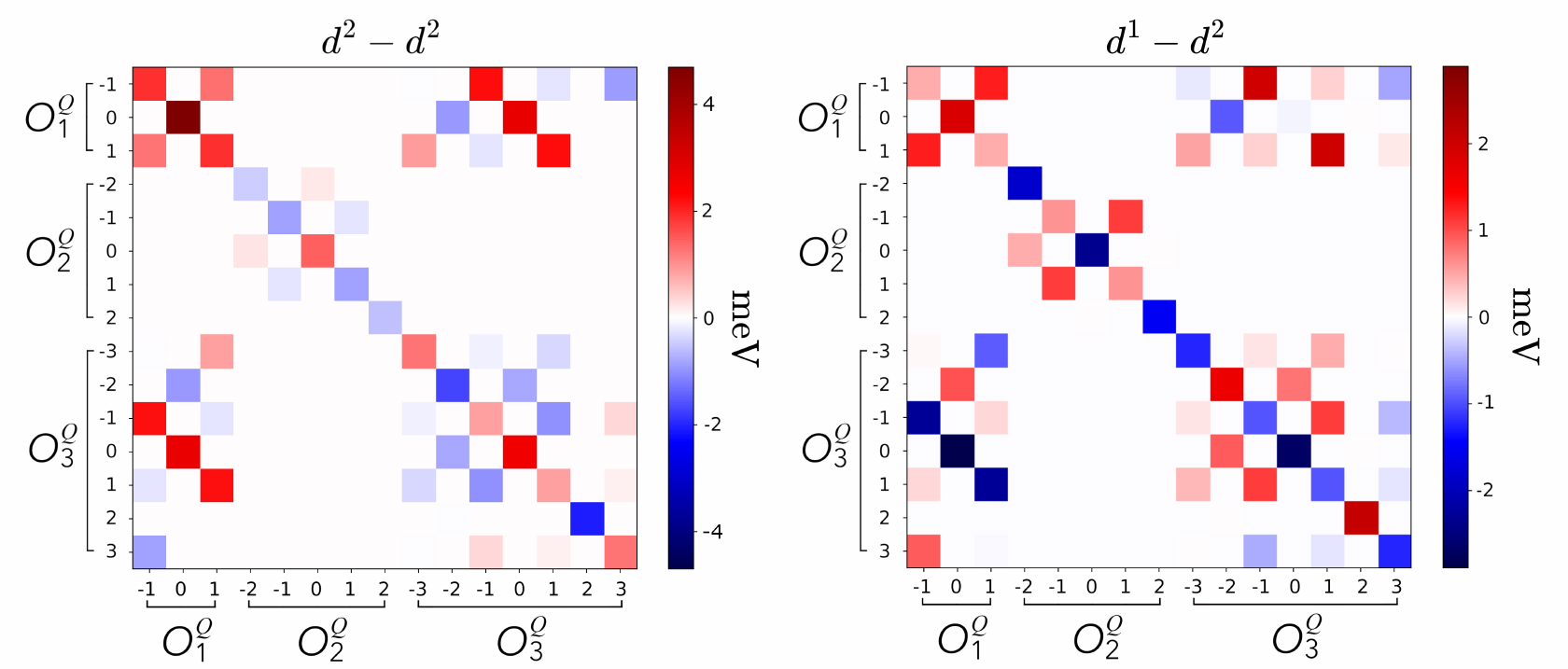}
    \caption{Color map of the IEI $V_{KK'}^{QQ'}$ in \bcnoo at pressure [0,1/2,0] Os-Os pair for (left) $d^2 -d^2$ interaction and (right) $d^1 -d^2$ interaction. All values are in meV. The numerical list of $V_{KK'}^{QQ'}$ is given in Suppl. Table~\ref{Tab:IEI1}.}
    \label{fig:iei1}
\end{figure}

In Suppl. Fig.~\ref{fig:iei1}, the IEI in \bcnoo are shown as a color map, with the corresponding values listed in Suppl. Table~\ref{Tab:IEI1} for the two cases of $d^2$–$d^2$ and $d^1$–$d^2$ electronic configurations. For the $d^1$–$d^2$ bond, both quadrupole–quadrupole and dipole–octupole interactions acquire antisymmetric components due to the inversion-symmetry breaking introduced by the polaron. The role of these antisymmetric interactions remains an open question and deserves further investigation.

  	\renewcommand\floatpagefraction{0.1}
  \begin{table}[!h]
  	\caption{\label{Tab:IEI1}  
  		 Calculated IEI $V_{KK’}^{QQ’}$ for the $J_{\text{eff}}=3/2$/$J_{\text{eff}}=2$ multiplets. The first two columns list $Q$ and $Q’$, while the third and fourth columns display the $KQ$ and $K’Q’$ tensors in Cartesian representation. The last three columns report the IEI values (in meV) for the [0, 1/2, 0] nearest-neighbor Os–Os bond in the supercell reference frame, comparing: (i) $d^2$–$d^2$ interactions from Ref.~\cite{Pourovskii2021}, (ii) $d^2$–$d^2$ interactions from this work, and (iii) $d^1$–$d^2$ interactions from this work. A slash denotes the reversed $Q/Q’$ case, highlighting the asymmetric contributions.
  	}
	\begin{center}
		\begin{ruledtabular}
			\renewcommand{\arraystretch}{1.2}
			\begin{tabular}{c c c c c c c}
\multicolumn{4}{c}{Bonds $\mathbf{R}_{ij}$ = [0.5, 0.5, 0] (primitive cell)} & $d^2 -d^2$ (Ref.~\cite{Pourovskii2021}) & $d^2 - d^2$ & $d^1 - d^2$ \\
		\hline
\multicolumn{7}{c}{Dipole-Dipole} \\
 				\hline
-1 & -1   & y  & y  &   1.62  &    1.89  &    0.45  \\
 0 &  0   & z  & z  &   4.17  &    4.70  &    1.87  \\
 1 & -1   & x  & y  &   1.26  &    1.28  &    1.28  \\
 1 &  1   & x  & x  &   1.62  &    1.89  &    0.45  \\
		\hline
		\hline
\multicolumn{7}{c}{Quadrupole-Quadrupole} \\
\hline
-2 & -2   & xy  & xy                    &   -0.41  &   -0.45  &   -1.84  \\
-1 & -1   & yz  & yz                    &   -0.79  &   -0.85  &    0.59 \\
 0 & -2   & z$^2$  & xy                 &    0.16  &    0.22  &    0.47/-0.01  \\
 0 &  0   & z$^2$  & z$^2$              &    1.32  &    1.45  &   -2.33  \\
 1 & -1   & xz  & yz                    &   -0.23  &   -0.24  &    1.09  \\
 1 &  1   & xz  & xz                    &   -0.79  &   -0.85  &    0.59  \\
 2 &  2   & x$^2$-y$^2$  & x$^2$-y$^2$  &   -0.58  &   -0.59  &   -1.52  \\
\hline
		\hline
\multicolumn{7}{c}{Octupole-Octupole} \\
\hline
-3 & -3   & y(x$^2$-3y$^2$)  & y(x$^2$-3y$^2$)  &    1.16  &    1.26  &   -1.24  \\
-2 & -2   & xyz  & xyz                          &   -1.49  &   -1.70  &    1.64  \\
-1 & -3   & yz$^2$  & y(x$^2$-3y$^2$)           &   -0.14  &   -0.13  &    0.16  \\
-1 & -1   & yz$^2$  & yz$^2$                    &    0.80  &    0.87  &   -0.95  \\
 0 & -2   & z$^3$  & xyz                        &   -0.79  &   -0.80  &    0.78  \\
 0 &  0   & z$^3$  & z$^3$                      &    2.35  &    2.58  &   -2.69  \\
 1 & -3   & xz$^2$  & y(x$^2$-3y$^2$)           &   -0.29  &   -0.35  &    0.39/0.46  \\
 1 & -1   & xz$^2$  & yz$^2$                    &   -0.98  &   -0.76  &    1.11  \\
 1 &  1   & xz$^2$  & xz$^2$                    &    0.80  &    0.86  &   -0.95  \\
 2 &  2   & z(x$^2$-y$^2$)  & z(x$^2$-y$^2$)    &   -1.89  &   -2.08  &    2.09  \\
 3 & -1   & x(3x$^2$-y$^2$)  & yz$^2$           &    0.29  &    0.35  &   -0.39  \\
 3 &  1   & x(3x$^2$-y$^2$)  & xz$^2$           &    0.14  &    0.12  &   -0.15  \\
 3 &  3   & x(3x$^2$-y$^2$)  & x(3x$^2$-y$^2$)  &    1.16  &    1.26  &   -1.24  \\
		\hline
		\hline
\multicolumn{7}{c}{Dipole-Octupole} \\
\hline
-1 & -3   & y  & y(x$^2$-3y$^2$)    &          &   -0.01  &   -0.13/0.04  \\
-1 & -1   & y  & yz$^2$             &    1.97  &    2.21  &    1.97/-2.27  \\
-1 &  1   & y  & xz$^2$             &   -0.20  &   -0.23  &    0.26/0.21  \\
-1 &  3   & y  & x(3x$^2$-y$^2$)    &   -0.89  &   -0.89  &   -0.51/0.91  \\
 0 & -2   & z  & xyz                &   -0.97  &   -0.95  &   -0.95/0.98  \\
 0 &  0   & z  & z$^3$              &    2.38  &    2.74  &   -0.07/-2.89  \\
 0 &  2   & z  & z(x$^2$-3y$^2$)    &          &    0.01  &          \\
 1 & -3   & x  & y(x$^2$-3y$^2$)    &    0.89  &    0.88  &    0.51/-0.91  \\
 1 & -1   & x  & yz$^2$             &   -0.20  &   -0.23  &    0.26/0.21  \\
 1 &  1   & x  & xz$^2$             &    1.97  &    2.21  &    1.98/-2.27  \\
 1 &  3   & x  & x(3x$^2$-y$^2$)    &          &    0.01  &    0.13/-0.04  \\
  			\end{tabular}
\end{ruledtabular}
\end{center}
\end{table}

\clearpage

\subsection{Wavefunctions and Crystal Fields}
In the following, we report the crystal-field splittings and wavefunctions of the ground-state multiplets, expressed in the total angular momentum $J=3/2$ and $J=2$ basis. One can observe that the tetragonal symmetry of the chosen supercell, together with the broken symmetry brought by the polaron localization, brings an equivalence between in-plane and out-of-plane osmium sites respectively, a feature that subsequently manifests itself in the MF results (see Fig.~\ref{fig:mf1}).

\vspace{1cm}

\noindent
\begin{minipage}[t]{0.48\linewidth}
\small\setlength{\tabcolsep}{6pt}\renewcommand{\arraystretch}{1.1}

\begin{tabular}{@{}lcccccc@{}}
\multicolumn{6}{c}{Energy (meV) for Os-1} \\
\hline
  & 0 & 0.66 &  &  &  \\
\hline
$\lvert 3/2,-3/2\rangle$ & $1/\sqrt{2}$    &  &       &       &  \\
$\lvert 3/2,-1/2\rangle$ &     &   $1/\sqrt{2}$    &  &    &       \\
$\lvert 3/3, 1/2\rangle$ &     &   $1/\sqrt{2}$    &       &       &       \\
$\lvert 3/2, 3/2\rangle$ &  $1/\sqrt{2}$   &       &  &  &       \\
\hline
\end{tabular}\\[0.8em]

\begin{tabular}{@{}lcccccc@{}}
\multicolumn{6}{c}{Energy (meV) for Os-2} \\
\hline
   & 0 & 2.3 & 19.3 & 19.3 & 21.4 \\
\hline
$\lvert 2,-2\rangle$ &     & $1/\sqrt{2}$ &       &       & $1/\sqrt{2}$ \\
$\lvert 2,-1\rangle$ &     &       & $1/\sqrt{2}$ & -$1/\sqrt{2}$&       \\
$\lvert 2, 0\rangle$ & 1   &       &       &       &       \\
$\lvert 2, 1\rangle$ &     &       & $1/\sqrt{2}$ & $1/\sqrt{2}$ &       \\
$\lvert 2, 2\rangle$ &     & $1/\sqrt{2}$ &       &       & -$1/\sqrt{2}$\\
\hline
\end{tabular}\\[0.8em]

\begin{tabular}{@{}lcccccc@{}}
\multicolumn{6}{c}{Energy (meV) for Os-3} \\
\hline
   & 0 & 5.2 & 19.1 & 22.3 & 23.7 \\
\hline
$\lvert 2,-2\rangle$ & -0.262    & 0.656 &       &       & $1/\sqrt{2}$ \\
$\lvert 2,-1\rangle$ &     &       & -$1/\sqrt{2}$ & $1/\sqrt{2}$&       \\
$\lvert 2, 0\rangle$ & 0.929   &   0.370    &       &       &       \\
$\lvert 2, 1\rangle$ &     &       & $1/\sqrt{2}$ & $1/\sqrt{2}$ &       \\
$\lvert 2, 2\rangle$ & -0.262   & 0.656 &       &       & -$1/\sqrt{2}$\\
\hline
\end{tabular}\\[0.8em]

\begin{tabular}{@{}lcccccc@{}}
\multicolumn{6}{c}{Energy (meV) for Os-4} \\
\hline
   & 0 & 5.2 & 19.1 & 22.3 & 23.7 \\
\hline
$\lvert 2,-2\rangle$ & -0.262    & 0.656 &       &       & $1/\sqrt{2}$ \\
$\lvert 2,-1\rangle$ &     &       & -$1/\sqrt{2}$ & $1/\sqrt{2}$&       \\
$\lvert 2, 0\rangle$ & 0.929   &   0.370    &       &       &       \\
$\lvert 2, 1\rangle$ &     &       & $1/\sqrt{2}$ & $1/\sqrt{2}$ &       \\
$\lvert 2, 2\rangle$ & -0.262   & 0.656 &       &       & -$1/\sqrt{2}$\\
\hline
\end{tabular}\\[0.8em]

\end{minipage}\hfill
\begin{minipage}[t]{0.48\linewidth}
\small\setlength{\tabcolsep}{6pt}\renewcommand{\arraystretch}{1.1}

\begin{tabular}{@{}lcccccc@{}}
\multicolumn{6}{c}{Energy (meV) for Os-5} \\
\hline
   & 0 & 5.2 & 19.1 & 22.3 & 23.7 \\
\hline
$\lvert 2,-2\rangle$ & -0.262    & 0.656 &       &       & $1/\sqrt{2}$ \\
$\lvert 2,-1\rangle$ &     &       & -$1/\sqrt{2}$ & $1/\sqrt{2}$&       \\
$\lvert 2, 0\rangle$ & 0.929   &   0.370    &       &       &       \\
$\lvert 2, 1\rangle$ &     &       & $1/\sqrt{2}$ & $1/\sqrt{2}$ &       \\
$\lvert 2, 2\rangle$ & -0.262   & 0.656 &       &       & -$1/\sqrt{2}$\\
\hline
\end{tabular}\\[0.8em]

\begin{tabular}{@{}lcccccc@{}}
\multicolumn{6}{c}{Energy (meV) for Os-6} \\
\hline
   & 0 & 5.2 & 19.1 & 22.3 & 23.7 \\
\hline
$\lvert 2,-2\rangle$ & -0.262    & 0.656 &       &       & $1/\sqrt{2}$ \\
$\lvert 2,-1\rangle$ &     &       & -$1/\sqrt{2}$ & $1/\sqrt{2}$&       \\
$\lvert 2, 0\rangle$ & 0.929   &   0.370    &       &       &       \\
$\lvert 2, 1\rangle$ &     &       & $1/\sqrt{2}$ & $1/\sqrt{2}$ &       \\
$\lvert 2, 2\rangle$ & -0.262   & 0.656 &       &       & -$1/\sqrt{2}$\\
\hline
\end{tabular}\\[0.8em]

\begin{tabular}{@{}lcccccc@{}}
\multicolumn{6}{c}{Energy (meV) for Os-7} \\
\hline
   & 0 & 1.0 & 19.0 & 19.9 & 19.9 \\
\hline
$\lvert 2,-2\rangle$ &$1/\sqrt{2}$    &  &       &       & $1/\sqrt{2}$ \\
$\lvert 2,-1\rangle$ &     &       & -$1/\sqrt{2}$ & $1/\sqrt{2}$&       \\
$\lvert 2, 0\rangle$ &    &   1    &       &       &       \\
$\lvert 2, 1\rangle$ &     &       & $1/\sqrt{2}$ & $1/\sqrt{2}$ &       \\
$\lvert 2, 2\rangle$ & $1/\sqrt{2}$   &  &       &       & -$1/\sqrt{2}$\\
\hline
\end{tabular}\\[0.8em]

\begin{tabular}{@{}lcccccc@{}}
\multicolumn{6}{c}{Energy (meV) for Os-8} \\
\hline
   & 0 & 1.0 & 19.0 & 19.9 & 19.9 \\
\hline
$\lvert 2,-2\rangle$ &$1/\sqrt{2}$    &  &       &       & $1/\sqrt{2}$ \\
$\lvert 2,-1\rangle$ &     &       & -$1/\sqrt{2}$ & $1/\sqrt{2}$&       \\
$\lvert 2, 0\rangle$ &    &   1    &       &       &       \\
$\lvert 2, 1\rangle$ &     &       & $1/\sqrt{2}$ & $1/\sqrt{2}$ &       \\
$\lvert 2, 2\rangle$ & $1/\sqrt{2}$   &  &       &       & -$1/\sqrt{2}$\\
\hline
\end{tabular}\\[0.8em]

\end{minipage}

\subsection{Ordered Phase and Mean Field values with temperature}

\begin{figure}[!h]
    \centering
    \includegraphics[width=0.9\textwidth]{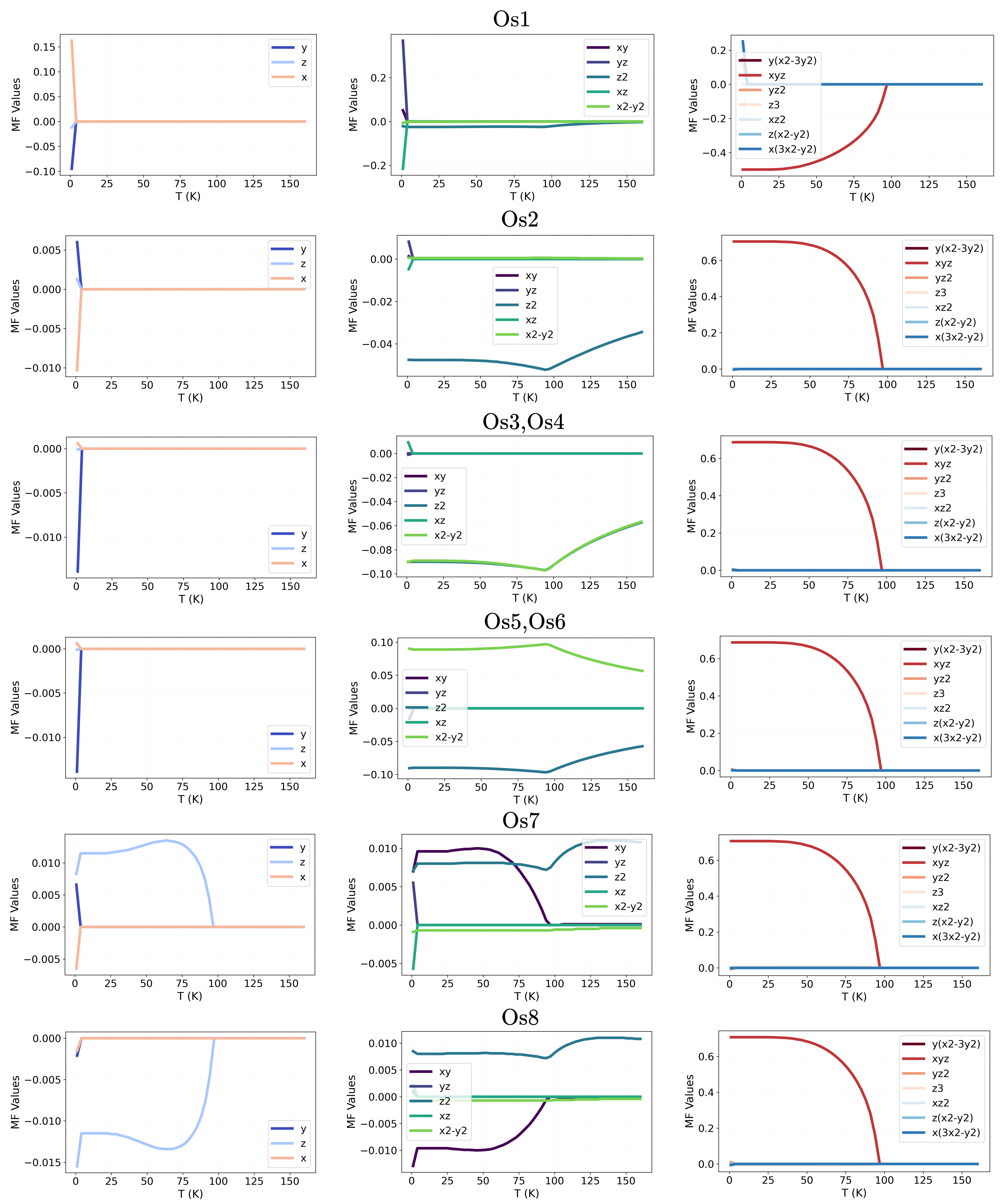}
    \caption{MF values of multipolar operators $O_K^{Q}$ expressed in Cartesian coordinates as a function of temperature for the different osmium ions of Fig.~\ref{fig:struct1}.}
    \label{fig:mf1}
\end{figure}

In Fig.~\ref{fig:mf1}, we show the MF values of different multipolar operators expressed in Cartesian coordinates. A sign reversal of the dominant octupolar exchange interaction is observed at $T_N \approx 94$ K. At higher temperatures, the only nonzero MF components are the quadrupoles with $z^2$ and $x^2-y^2$ character. Their finite values arise from the combined effects of the polaronic charge and the tetragonal crystal field imposed by the supercell geometry, though their magnitude remains very small ($\sim 0.1$). At lower temperatures ($\sim 4$ K), a second transition appears, involving a more complex ordering of dipoles and quadrupoles. We do not discuss this transition in the Main text, as it is most likely an artifact of volumetric constraints in the supercell; notably, it vanishes once the polaronic structure is fully relaxed.

\clearpage

\subsection{Intersite Exchange Interactions in spin+orbital t$_{2g}$ basis}

In the following, the non-SOC $t_{2g}$-IEI of \bcnoo  in the spin+orbital basis are are listed. The $t_{2g}$-IEI refer to the $I^{\alpha \alpha', \beta \beta'}_{s s', l l'} (ij)$ of equation (2) in the main.

  	\renewcommand\floatpagefraction{0.1}
  \begin{table}[!h]
  	\caption{\label{Tab:IEI1}  
  		 Calculated IEI $I^{\alpha \alpha', \beta \beta'}_{s s', l l'} (ij)$ for the  spin+orbital $t_{2g}$ basis. The first two columns list spin-$s$ component ($\alpha$) and orbital-$l$ component $\beta$ on the first site, while the other two list spin-$s'$ component ($\alpha'$) and orbital-$l'$ component $\beta'$ of the second site. The last two columns report the IEI values (in meV) for the [0, 1/2, 0] nearest-neighbor Os–Os bond in the supercell reference frame, comparing: (i) $d^2$–$d^2$ interactions and (ii) $d^1$–$d^2$ interactions. The values highlighted in bold are the ones who map onto $xyz$ $J_{\mathrm{eff}}$-IEI. 
  	}
	\begin{center}
		\begin{ruledtabular}
			\renewcommand{\arraystretch}{1.2}
			\begin{tabular}{c c c c c c }
\multicolumn{4}{c}{Bonds $\mathbf{R}_{ij}$ = [0.5, 0.5, 0] (primitive cell)} & $d^2 - d^2$ & $d^1 - d^2$ \\
		\hline
\multicolumn{6}{c}{$(s;l) - (s';l')$} \\
		\hline
        		\hline
\multicolumn{6}{c}{(Monopole;Dipole)-(Monopole;Dipole)} \\
 				\hline
 1 & y    & 1  & y  &  -4.42    &    -4.24  \\
 1 & y    & 1  & x  &   2.78   &    2.92  \\
 1 & z    & 1  & z  &  0.92    &    0.76  \\
 1 & x    & 1  & x  &  -4.42    &    -4.24  \\
		\hline
		\hline
\multicolumn{6}{c}{(Monopole;Dipole)-(Monopole;Quadrupole)} \\
 				\hline
 1 & y    & 1  & xy      &   -0.04   &  -0.06   \\
 1 & y    & 1  & z$^2$   &   0.03   &  -0.04   \\
 1 & x    & 1  & xy      &   0.04   &  0.06   \\
 1 & x    & 1  & z$^2$   &   -0.03   &  0.04   \\
		\hline
		\hline
\multicolumn{6}{c}{(Monopole;Quadrupole)-(Monopole;Quadrupole)} \\
 				\hline
 1 & xy    & 1  & xy      &  2.31    &  -2.21   \\
 1 & xy    & 1  & z$^2$   &  0.87    &  -1.39   \\
 1 & yz    & 1  & yz      & -4.72     &  4.55   \\
 1 & yz    & 1  & xz      & -3.03     &  3.19   \\
 1 & z$^2$ & 1  & z$^2$   & 10.07    &  -10.05   \\
 1 & xz    & 1  & xz      &  -4.72    &  4.55   \\
 1 & x$^2$-y$^2$    & 1  & x$^2$-y$^2$      &  0.97    &  -0.78   \\
		\hline
		\hline
\multicolumn{6}{c}{(Dipole;Monopole)-(Dipole;Monopole)} \\
 				\hline
 y & 1    & y  & 1      & 10.83     &  5.75   \\
 z & 1    & z  & 1      & 10.83     &  5.75   \\
 x & 1    & x  & 1      & 10.83     &  5.75   \\
		\hline
		\hline
\multicolumn{6}{c}{(Dipole;Monopole)-(Dipole;Dipole)} \\
 				\hline
 y & 1    & y  & y      &  -0.11    &  0.04   \\
 y & 1    & y  & x      &  0.11    &   -0.04   \\
 z & 1    & z  & y      &  -0.11    &   0.04   \\
 z & 1    & z  & x      &  0.11    &  -0.04   \\
 x & 1    & x  & y      &  -0.11    &   0.04   \\
 x & 1    & x  & x      &  0.11    &  -0.04   \\
		\hline
		\hline
\multicolumn{6}{c}{(Dipole;Monopole)-(Dipole;Quadrupole)} \\
 				\hline
 y & 1    & y  & xy         & -1.85     &   -0.93  \\
 y & 1    & y  & z$^2$      &  5.65    &   2.87   \\
 z & 1    & z  & xy         & -1.85     &  -0.93   \\
 z & 1    & z  & z$^2$      &  5.65    &   2.87   \\
 x & 1    & x  & xy         & -1.85     &  -0.93   \\
 x & 1    & x  & z$^2$      &  5.65    &  2.87   \\
  			\end{tabular}
\end{ruledtabular}
\end{center}
\end{table}

  	\renewcommand\floatpagefraction{0.1}
  \begin{table}[!h]
	\begin{center}
		\begin{ruledtabular}
			\renewcommand{\arraystretch}{1.2}
			\begin{tabular}{c c c c c c }
\multicolumn{4}{c}{Bonds $\mathbf{R}_{ij}$ = [0.5, 0.5, 0] (primitive cell)} & $d^2 - d^2$ & $d^1 - d^2$ \\
		\hline
\multicolumn{6}{c}{$(s;l) - (s';l')$} \\
		\hline
        		\hline
\multicolumn{6}{c}{(Dipole;Dipole)-(Dipole;Dipole)} \\
 				\hline
 y & y    & y  & y  &  -2.82    &   -3.24    \\
 y & y    & y  & x  &   1.74   &    2.18   \\
 y & z    & y  & z  &   0.60   &    0.60   \\
 y & x    & y  & x  &   -2.82   &   -3.24   \\
 z & y    & z  & y  &   -2.82   &   -3.24    \\
 z & y    & z  & x  &   1.74   &    2.18   \\
 z & z    & z  & z  &   0.60   &    0.60   \\
 z & x    & z  & x  &  -2.82    &   -3.24    \\
 x & y    & x  & y  &  -2.82   &   -3.24    \\
 x & y    & x  & x  &   1.74   &    2.18   \\
 x & z    & x  & z  &   0.60   &    0.60   \\
 x & x    & x  & x  &  -2.82    &   -3.24    \\
		\hline
		\hline
\multicolumn{6}{c}{(Dipole;Dipole)-(Dipole;Quadrupole)} \\
 				\hline
 y & y    & y  & xy  &  -0.03    &   -0.05    \\
 y & x    & y  & xy  &   0.02   &    0.05   \\
 z & y    & z  & xy  &  -0.03    &   -0.05    \\
 z & x    & z  & xy  &   0.02   &    0.05    \\
 x & y    & x  & xy  &  -0.03    &   -0.05    \\
 x & x    & x  & xy  &   0.02   &    0.05    \\
		\hline
		\hline
\multicolumn{6}{c}{(Dipole;Quadrupole)-(Dipole;Quadrupole)} \\
 				\hline
 y & xy    & y  & xy                        &  1.23    &  -1.39    \\
 y & xy    & y  & z$^2$                     &  0.39    &   -0.78   \\
 y & yz    & y  & yz                        & -2.62    &   2.98   \\
 y & yz    & y  & xz                        & -1.59    &   1.96   \\
 y & z$^2$    & y  & z$^2$                  &  5.58    &  -6.53    \\
\textbf{y} & \textbf{xz}    & \textbf{y}  & \textbf{xz}  &  \textbf{-2.62}  &  \textbf{2.98}   \\
 y & x$^2$-y$^2$ & y  & x$^2$-y$^2$         &  0.57    &   -0.56   \\
\textbf{z} & \textbf{xy}    & \textbf{z}  & \textbf{xy}   &  \textbf{1.23}    &   \textbf{-1.39}    \\
 z & xy    & z  & z$^2$                     &  0.39    &   -0.78   \\
 z & yz    & z  & yz                        &  -2.62    &   2.98   \\
 z & yz    & z  & xz                        & -1.59     &   1.96   \\
 z & z$^2$    & z  & z$^2$                  & 5.58     &   -6.53   \\
 z & xz    & z  & xz                        &  -2.62    &   2.98   \\
 z & x$^2$-y$^2$ & z  & x$^2$-y$^2$         &  0.57    &   -0.56   \\
x & xy     &  x & xy                        &  1.23    &  -1.39    \\
 x & xy    & x  & z$^2$                     &   0.39   &   -0.78   \\
\textbf{x} & \textbf{yz}    & \textbf{x}  & \textbf{yz} & \textbf{-2.62}     &   \textbf{2.98}   \\
 x & yz    & x  & xz                        &  -1.59    &   1.96   \\
 x & z$^2$    & x  & z$^2$                  &  5.58    &  -6.53    \\
 x & xz    & x  & xz                        &  -2.62    &   2.98   \\
 x & x$^2$-y$^2$ & x  & x$^2$-y$^2$         &   0.57   &  -0.56    \\
  			\end{tabular}
\end{ruledtabular}
\end{center}
\end{table}

\clearpage

\section{Doping concentration $\delta = 0.125$ from constrained DFT+U structure}

In the following, we present DFT+HI, FT-HI, and mean-field (MF) results for \bcnoo within the virtual crystal approximation at Na doping concentration $\delta = 0.125$ and with the polaronic distorted structure obtained from constrained DFT+U calculation. Section A describes the structural details, Section B reports the DFT+HI density of states, Section C discusses the $J_{\mathrm{eff}}$-IEI, Section D analyzes the crystal-field splittings and ground-state multiplet wavefunctions, and Section E presents the ordered phase and the MF order parameters as a function of temperature.  

\subsection{Structure}

\begin{figure}[!h]
    \centering
    \includegraphics[width=0.45\textwidth]{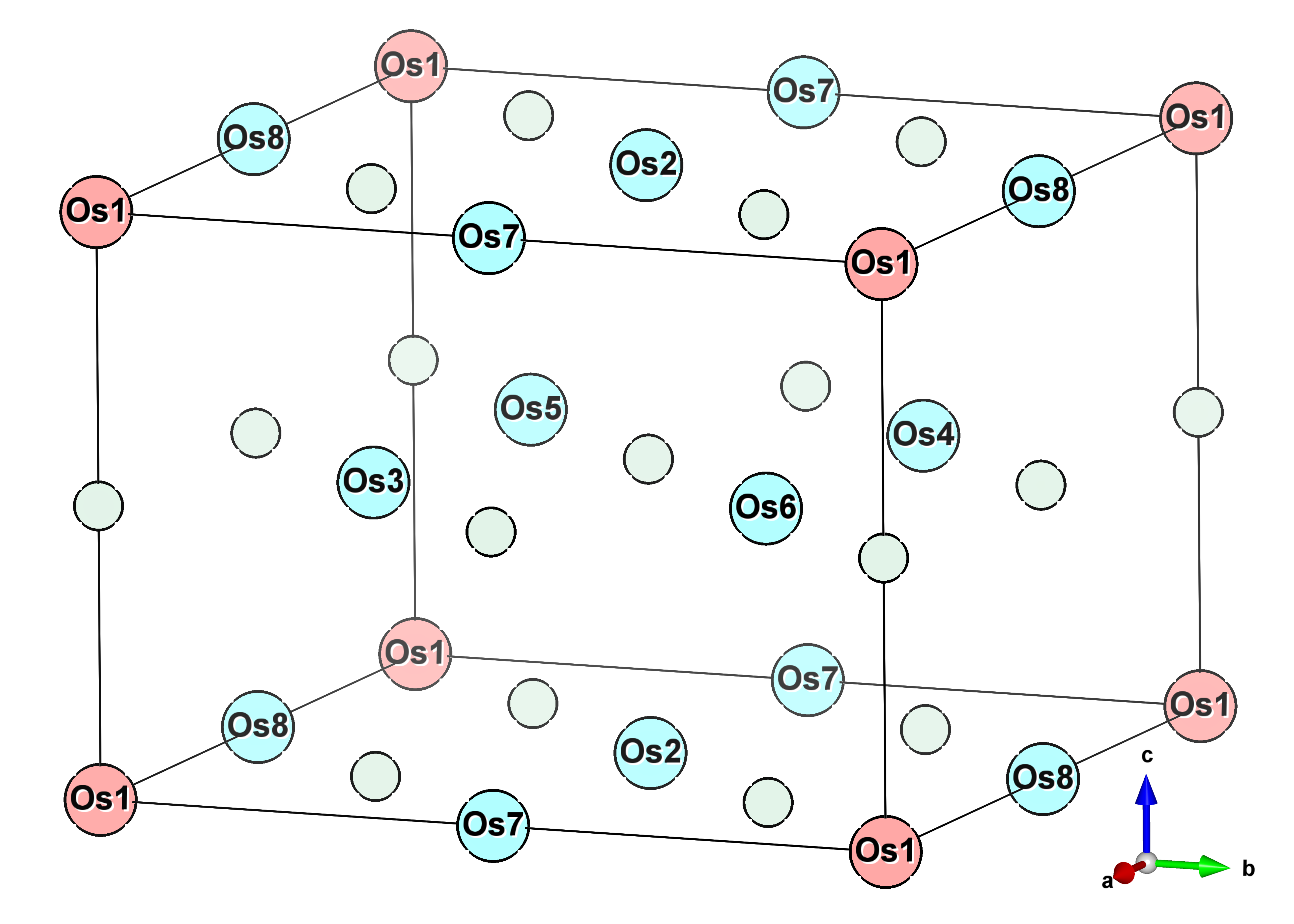}
    \caption{Supercell of \bcnoo with labels for the osmium ions as indexed in the next Sections. The polaron is localized on the Os1 (light red).}
    \label{fig:struct2}
\end{figure}

\subsection{Polaronic density of States in constrained DFT+U}

\begin{figure}[!h]
    \centering
    \includegraphics[width=0.45\textwidth]{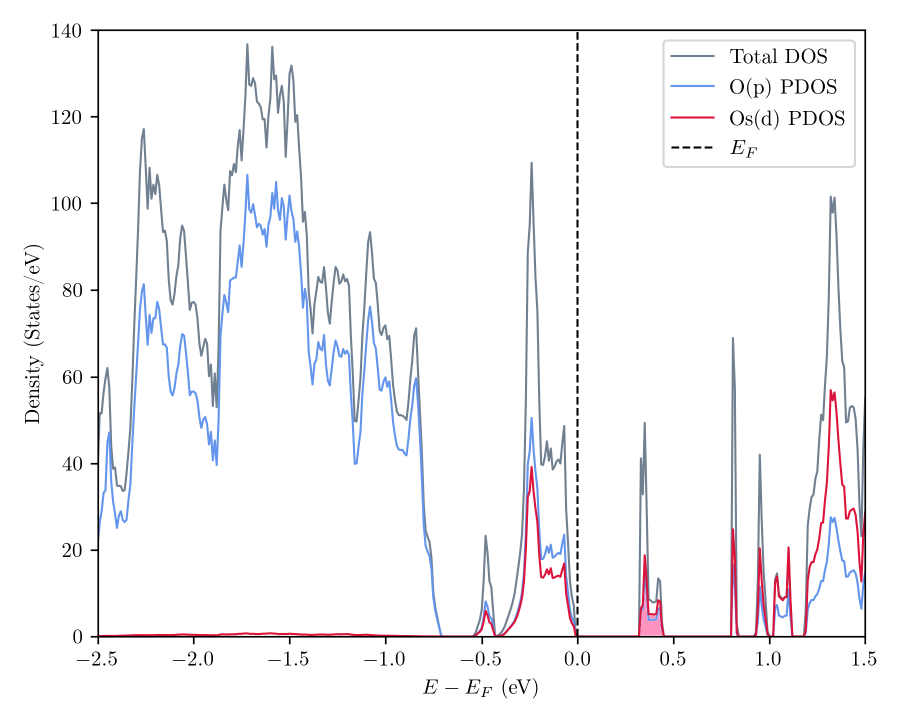}
    \caption{DOS from constrained DFT+U. The hole-polaron peak is highlighted in light-red.}
    \label{fig:dos2}
\end{figure}

\subsection{Intersite Exchange Interactions in J$_{\mathrm{eff}}$ basis}

\begin{figure}[!h]
    \centering
    \includegraphics[width=0.9\textwidth]{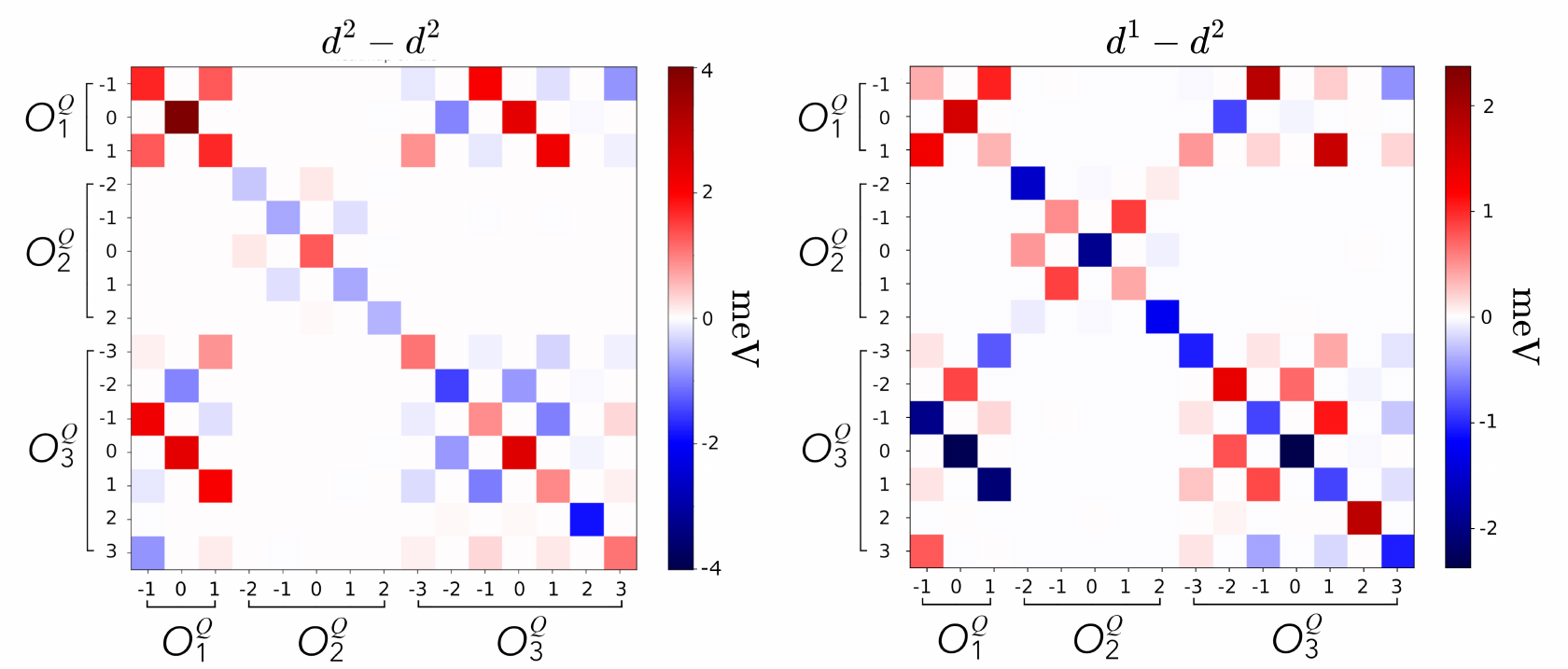}
    \caption{Color map of the IEI $V_{KK'}^{QQ'}$ in \bcnoo at pressure [0,1/2,0] Os-Os pair for (left) $d^2 -d^2$ interaction and (right) $d^1 -d^2$ interaction. All values are in meV. The numerical list of $V_{KK'}^{QQ'}$ is given in Suppl. Table~\ref{Tab:IEI2}.}
    \label{fig:iei2}
\end{figure}

In Suppl. Fig.~\ref{fig:iei2}, the IEI in \bcnoo are shown as a color map, with the corresponding values listed in Suppl. Table~\ref{Tab:IEI2} for the two cases of $d^2$–$d^2$ and $d^1$–$d^2$ electronic configurations. For the $d^1$–$d^2$ bond, an even stronger antisymmetric IEI matrix appears both in  quadrupole–quadrupole and dipole–octupole interactions induced by the polaron.

  	\renewcommand\floatpagefraction{0.1}
  \begin{table}[!h]
  	\caption{\label{Tab:IEI2}  
  		 Calculated IEI $V_{KK’}^{QQ’}$ for the $J_{\text{eff}}=3/2$/$J_{\text{eff}}=2$ multiplets. The first two columns list $Q$ and $Q’$, while the third and fourth columns display the $KQ$ and $K’Q’$ tensors in Cartesian representation. The last three columns report the IEI values (in meV) for the [0, 1/2, 0] nearest-neighbor Os–Os bond in the supercell reference frame, comparing: (i) $d^2$–$d^2$ interactions from Ref.~\cite{Pourovskii2021}, (ii) $d^2$–$d^2$ interactions from this work, and (iii) $d^1$–$d^2$ interactions from this work. A slash denotes the reversed $Q/Q’$ case, highlighting the asymmetric contributions.
  	}
	\begin{center}
		\begin{ruledtabular}
			\renewcommand{\arraystretch}{1.2}
			\begin{tabular}{c c c c c c c}
\multicolumn{4}{c}{Bonds $\mathbf{R}_{ij}$ = [0.5, 0.5, 0] (primitive cell)} & $d^2 -d^2$ (Ref.~\cite{Pourovskii2021}) & $d^2 - d^2$ & $d^1 - d^2$ \\
		\hline
\multicolumn{7}{c}{Dipole-Dipole} \\
 				\hline
-1 & -1   & y  & y  &   1.62  &    1.70  &    0.39  \\
 0 &  0   & z  & z  &   4.17  &    4.01  &    1.56  \\
 1 & -1   & x  & y  &   1.26  &    1.30  &    1.03  \\
 1 &  1   & x  & x  &   1.62  &    1.68  &    0.35  \\
		\hline
		\hline
\multicolumn{7}{c}{Quadrupole-Quadrupole} \\
\hline
-2 & -2   & xy  & xy                    &   -0.41  &   -0.42  &   -1.52  \\
-2 &  2   & xy  & x$^2$-y$^2$           &          &          &    0.09/-0.09  \\
-1 & -1   & yz  & yz                    &   -0.79  &   -0.69  &    0.53 \\
 0 & -2   & z$^2$  & xy                 &    0.16  &    0.17  &    0.48/-0.02  \\
 0 &  0   & z$^2$  & z$^2$              &    1.32  &    1.29  &   -1.92  \\
 1 & -1   & xz  & yz                    &   -0.23  &   -0.23  &    0.90  \\
 1 &  1   & xz  & xz                    &   -0.79  &   -0.68  &    0.40  \\
 2 &  2   & x$^2$-y$^2$  & x$^2$-y$^2$  &   -0.58  &   -0.56  &   -1.27  \\
\hline
		\hline
\multicolumn{7}{c}{Octupole-Octupole} \\
\hline
-3 & -3   & y(x$^2$-3y$^2$)  & y(x$^2$-3y$^2$)  &    1.16  &    1.10  &   -1.05  \\
-2 & -2   & xyz  & xyz                          &   -1.49  &   -1.46  &    1.39  \\
-1 & -3   & yz$^2$  & y(x$^2$-3y$^2$)           &   -0.14  &   -0.11  &    0.12  \\
-1 & -1   & yz$^2$  & yz$^2$                    &    0.80  &    0.89  &   -0.86  \\
 0 & -2   & z$^3$  & xyz                        &   -0.79  &   -0.80  &    0.8/0.7  \\
 0 &  0   & z$^3$  & z$^3$                      &    2.35  &    2.51  &   -2.37  \\
 1 & -3   & xz$^2$  & y(x$^2$-3y$^2$)           &   -0.29  &   -0.31  &    0.27/0.40  \\
 1 & -1   & xz$^2$  & yz$^2$                    &   -0.98  &   -0.99  &    1.08  \\
 1 &  1   & xz$^2$  & xz$^2$                    &    0.80  &    0.92  &   -0.87  \\
 2 &  2   & z(x$^2$-y$^2$)  & z(x$^2$-y$^2$)    &   -1.89  &   -1.87  &    1.78  \\
 3 & -1   & x(3x$^2$-y$^2$)  & yz$^2$           &    0.29  &    0.30  &   -0.25  \\
 3 &  1   & x(3x$^2$-y$^2$)  & xz$^2$           &    0.14  &    0.12  &   -0.14  \\
 3 &  3   & x(3x$^2$-y$^2$)  & x(3x$^2$-y$^2$)  &    1.16  &    1.09  &   -1.05  \\
		\hline
		\hline
\multicolumn{7}{c}{Dipole-Octupole} \\
\hline
-1 & -3   & y  & y(x$^2$-3y$^2$)    &          &    0.10  &   -0.03/0.12  \\
-1 & -1   & y  & yz$^2$             &    1.97  &    2.21  &    1.82/-1.96  \\
-1 &  1   & y  & xz$^2$             &   -0.20  &   -0.16  &    0.26/0.12  \\
-1 &  3   & y  & x(3x$^2$-y$^2$)    &   -0.89  &   -0.84  &   -0.51/0.77  \\
 0 & -2   & z  & xyz                &   -0.97  &   -0.95  &   -0.85/0.87  \\
 0 &  0   & z  & z$^3$              &    2.38  &    2.41  &   -0.04/-2.31  \\
 0 &  2   & z  & z(x$^2$-3y$^2$)    &          &    0.03  &    0.01/0.01        \\
 1 & -3   & x  & y(x$^2$-3y$^2$)    &    0.89  &    0.82  &    0.47/-0.77  \\
 1 & -1   & x  & yz$^2$             &   -0.20  &   -0.24  &    0.20/0.17  \\
 1 &  1   & x  & xz$^2$             &    1.97  &    2.10  &    1.68/-2.09  \\
 1 &  3   & x  & x(3x$^2$-y$^2$)    &          &    0.15  &    0.20/-0.01  \\
  			\end{tabular}
\end{ruledtabular}
\end{center}
\end{table}

\clearpage

\subsection{Wavefunctions and Crystal Fields}
In the following, we report the crystal-field splittings and wavefunctions of the ground-state multiplets, expressed in the total angular momentum $J=3/2$ and $J=2$ basis. One can observe that the relaxed geometry  of the constrained  supercell, in this case, together with the broken symmetry brought by the polaron localization, breaks the equivalence between in-plane and out-of-plane osmium sites as observed previously. This result impact on the MF calculation (see Fig.~\ref{fig:mf2}).

\vspace{1cm}

\noindent
\begin{minipage}[t]{0.48\linewidth}
\small\setlength{\tabcolsep}{6pt}\renewcommand{\arraystretch}{1.1}

\begin{tabular}{@{}lcccccc@{}}
\multicolumn{6}{c}{Energy (meV) for Os-1} \\
\hline
  & 0 & 7.2 &  &  &  \\
\hline
$\lvert 3/2,-3/2\rangle$ & $1/\sqrt{2}$    &  &       &       &  \\
$\lvert 3/2,-1/2\rangle$ &     &   $1/\sqrt{2}$    &  &    &       \\
$\lvert 3/3, 1/2\rangle$ &     &   $1/\sqrt{2}$    &       &       &       \\
$\lvert 3/2, 3/2\rangle$ &  $1/\sqrt{2}$   &       &  &  &       \\
\hline
\end{tabular}\\[0.8em]

\begin{tabular}{@{}lcccccc@{}}
\multicolumn{6}{c}{Energy (meV) for Os-2} \\
\hline
   & 0 & 10.2 & 19.6 & 20.6 & 28.6 \\
\hline
$\lvert 2,-2\rangle$ & 0.590    & 0.388 &      &  $1/\sqrt{2}$      &  \\
$\lvert 2,-1\rangle$ &          &       & -$1/\sqrt{2}$  &    &   $1/\sqrt{2}$     \\
$\lvert 2, 0\rangle$ & -0.550   & 0.835 &       &       &       \\
$\lvert 2, 1\rangle$ &          &       &  $1/\sqrt{2}$  &   &  $1/\sqrt{2}$     \\
$\lvert 2, 2\rangle$ &  0.590   & 0.388 &      &  -$1/\sqrt{2}$      & \\
\hline
\end{tabular}\\[0.8em]

\begin{tabular}{@{}lcccccc@{}}
\multicolumn{6}{c}{Energy (meV) for Os-3} \\
\hline
   & 0 & 17.9 & 20.2 & 32.7 & 34.4 \\
\hline
$\lvert 2,-2\rangle$ & $1/\sqrt{2}$   &$1/\sqrt{2}$  &       &       & \\
$\lvert 2,-1\rangle$ &     &       &  & $1/\sqrt{2}$&   $1/\sqrt{2}$    \\
$\lvert 2, 0\rangle$ &     &        &   1    &       &       \\
$\lvert 2, 1\rangle$ &     &       &  & $1/\sqrt{2}$ &   -$1/\sqrt{2}$    \\
$\lvert 2, 2\rangle$ & $1/\sqrt{2}$   & -$1/\sqrt{2}$  &       &       &  \\
\hline
\end{tabular}\\[0.8em]

\begin{tabular}{@{}lcccccc@{}}
\multicolumn{6}{c}{Energy (meV) for Os-4} \\
\hline
   & 0 & 1.7 & 13.7 & 14.2 & 15.9 \\
\hline
$\lvert 2,-2\rangle$ &     & $1/\sqrt{2}$ &       &       & $1/\sqrt{2}$ \\
$\lvert 2,-1\rangle$ &     &       & $1/\sqrt{2}$ & -$1/\sqrt{2}$&       \\
$\lvert 2, 0\rangle$ & 1   &      &       &       &       \\
$\lvert 2, 1\rangle$ &     &       & $1/\sqrt{2}$ & $1/\sqrt{2}$ &       \\
$\lvert 2, 2\rangle$ &     & $1/\sqrt{2}$ &       &       & -$1/\sqrt{2}$\\
\hline
\end{tabular}\\[0.8em]

\end{minipage}\hfill
\begin{minipage}[t]{0.48\linewidth}
\small\setlength{\tabcolsep}{6pt}\renewcommand{\arraystretch}{1.1}

\begin{tabular}{@{}lcccccc@{}}
\multicolumn{6}{c}{Energy (meV) for Os-5} \\
\hline
   & 0 & 4.7 & 18.5 & 19.6 & 22.4 \\
\hline
$\lvert 2,-2\rangle$ & 0.655    & -0.266 &  $1/\sqrt{2}$     &       &  \\
$\lvert 2,-1\rangle$ &     &       &   & $1/\sqrt{2}$&   $1/\sqrt{2}$    \\
$\lvert 2, 0\rangle$ & 0.375   &   0.926    &       &       &       \\
$\lvert 2, 1\rangle$ &     &       &  & $1/\sqrt{2}$ &  -$1/\sqrt{2}$     \\
$\lvert 2, 2\rangle$ & 0.655   & -0.266 &  -$1/\sqrt{2}$     &       & \\
\hline
\end{tabular}\\[0.8em]

\begin{tabular}{@{}lcccccc@{}}
\multicolumn{6}{c}{Energy (meV) for Os-6} \\
\hline
   & 0 & 3.4 & 18.5 & 18.8 & 21.1 \\
\hline
$\lvert 2,-2\rangle$ & 0.636    & -0.307 &  $1/\sqrt{2}$     &       &  \\
$\lvert 2,-1\rangle$ &     &       &  & $1/\sqrt{2}$&    $1/\sqrt{2}$   \\
$\lvert 2, 0\rangle$ & 0.435   &  0.900    &       &       &       \\
$\lvert 2, 1\rangle$ &     &       &   & $1/\sqrt{2}$ &  -$1/\sqrt{2}$     \\
$\lvert 2, 2\rangle$ & 0.636   & -0.307 & -$1/\sqrt{2}$     &       & \\
\hline
\end{tabular}\\[0.8em]

\begin{tabular}{@{}lcccccc@{}}
\multicolumn{6}{c}{Energy (meV) for Os-7} \\
\hline
   & 0 & 8.8 & 19.0 & 21.2 & 27.1 \\
\hline
$\lvert 2,-2\rangle$ & 0.648    & -0.282 &  $1/\sqrt{2}$     &       &  \\
$\lvert 2,-1\rangle$ &     &       &  & $1/\sqrt{2}$&    $1/\sqrt{2}$   \\
$\lvert 2, 0\rangle$ & 0.399   &  0.917    &       &       &       \\
$\lvert 2, 1\rangle$ &     &       &   & $1/\sqrt{2}$ &  -$1/\sqrt{2}$     \\
$\lvert 2, 2\rangle$ & 0.648   & -0.282 & -$1/\sqrt{2}$     &       & \\
\hline
\end{tabular}\\[0.8em]

\begin{tabular}{@{}lcccccc@{}}
\multicolumn{6}{c}{Energy (meV) for Os-7} \\
\hline
   & 0 & 10.5 & 20.0 & 20.8 & 29.1 \\
\hline
$\lvert 2,-2\rangle$ & 0.621    & -0.338 &  $1/\sqrt{2}$     &       &  \\
$\lvert 2,-1\rangle$ &     &       &  & $1/\sqrt{2}$&    $1/\sqrt{2}$   \\
$\lvert 2, 0\rangle$ & 0.478   &  0.878    &       &       &       \\
$\lvert 2, 1\rangle$ &     &       &   & $1/\sqrt{2}$ &  -$1/\sqrt{2}$     \\
$\lvert 2, 2\rangle$ & 0.621   & -0.338 & -$1/\sqrt{2}$     &       & \\
\hline
\end{tabular}\\[0.8em]

\end{minipage}

\subsection{Ordered Phase and Mean Field values with temperature}

\begin{figure}[!h]
    \centering
    \includegraphics[width=0.9\textwidth]{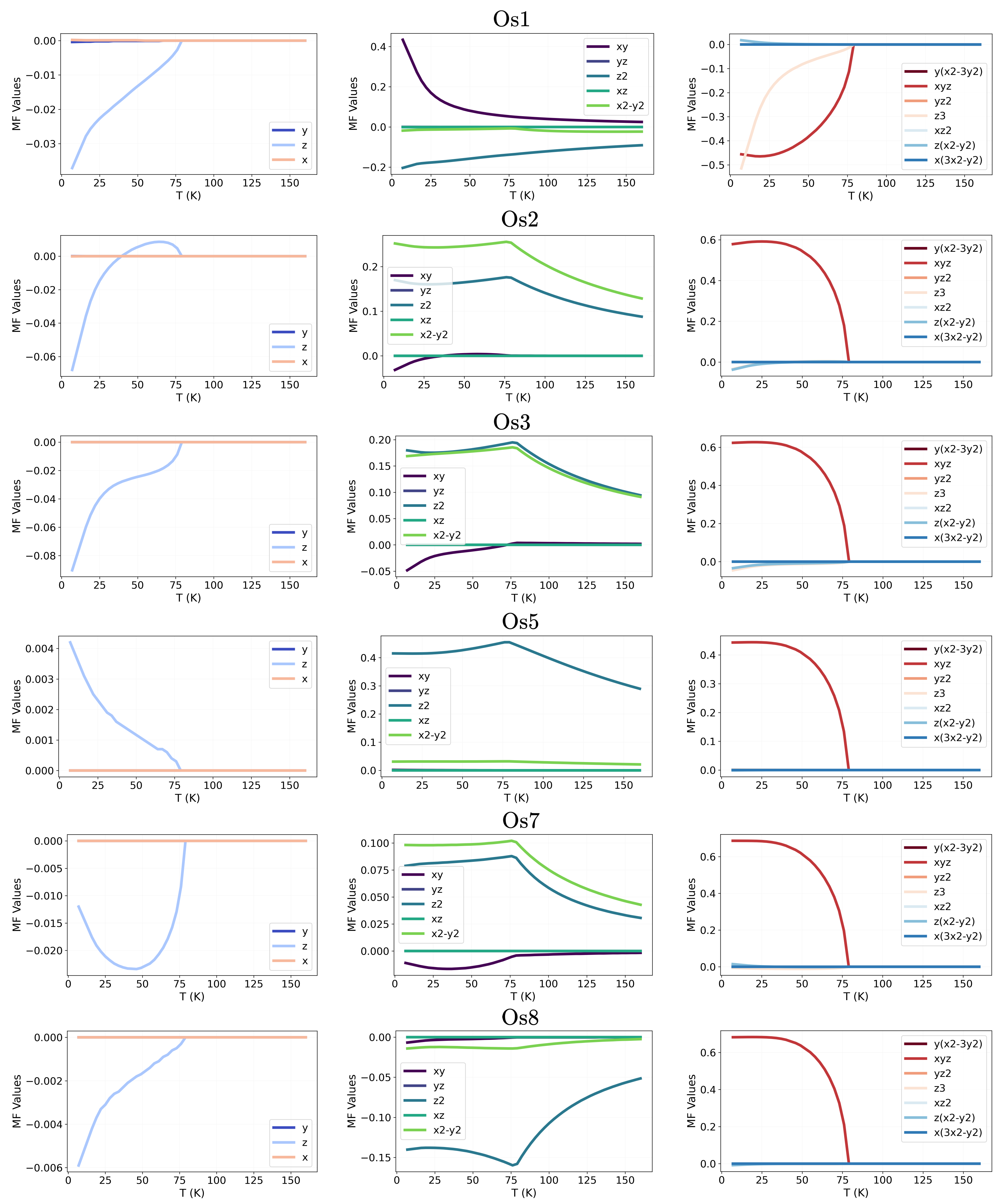}
    \caption{MF values of multipolar operators $O_K^{Q}$ expressed in Cartesian coordinates as a function of temperature for the different osmium ions of Fig.~\ref{fig:struct2}.}
    \label{fig:mf2}
\end{figure}

In Fig.\ref{fig:mf2}, we present the MF values of different multipolar operators in Cartesian coordinates. A sign reversal of the dominant octupolar exchange interaction occurs at $T_N \approx 77$ K. At higher temperatures, the only nonzero MF components are quadrupoles of $z^2$ and $x^2-y^2$ character, consistent with the undistorted case. Importantly, in line with the constrained DFT+U result, the supercell exhibits a weak ferromagnetic moment of $\sim 0.02 , \mu_B$. At low temperature, the ordered state involves the $z^3$ octupole, whose $\Gamma_2$ symmetry matches that of the dipoles ($z$). This behavior reflects a crystal-field effect: the presence of the polaron lowers nearby $T_{2g}$ states sufficiently to favor $z^3$ ordering, even though the change in IEI relative to the undistorted case is not particularly strong (see Tab.\ref{Tab:IEI1} and Tab.~\ref{Tab:IEI2}).

\clearpage

\section{Doping concentration $\delta = 0.125$ with chemical doping}

In the following, we present DFT+HI, FT-HI, and MF results for \bcnoo with chemical substitution of Ca with Na ions at doping concentration $\delta = 0.125$. Section A describes the structural details, Section B reports the DFT+HI density of states, Section C discusses the $J_{\mathrm{eff}}$-IEI, Section D analyzes the crystal-field splittings and ground-state multiplet wavefunctions, and Section E presents the ordered phase and the MF order parameters as a function of temperature.  

\subsection{Structure}

\begin{figure}[!h]
    \centering
    \includegraphics[width=0.45\textwidth]{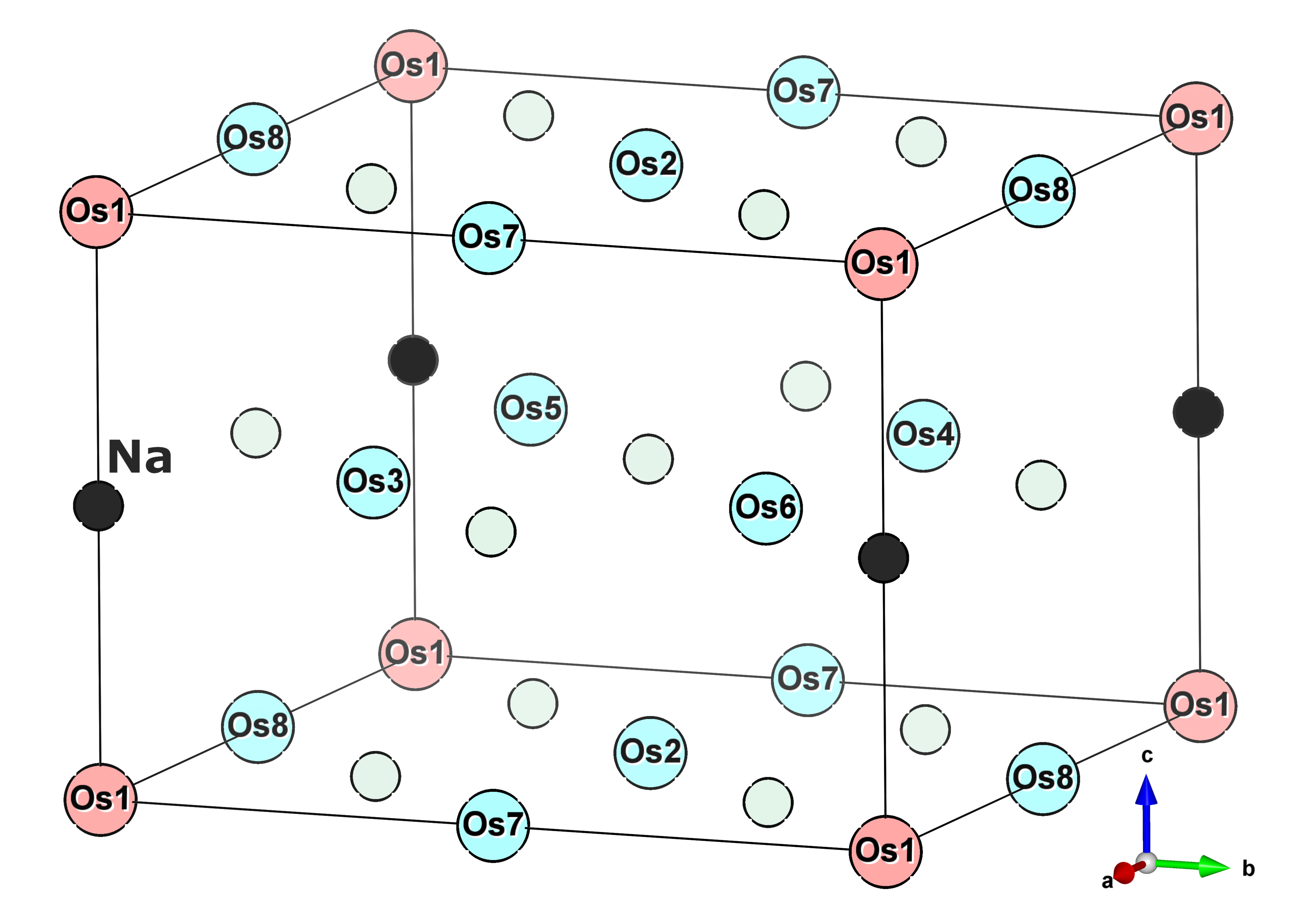}
    \caption{Supercell of \bcnoo with labels for the osmium ions as indexed in the next Sections. The polaron is localized on the Os1 (light red).}
    \label{fig:struct3}
\end{figure}

\subsection{Polaronic density of States in DFT, DFT+HI}

\begin{figure}[!h]
    \centering
    \includegraphics[width=0.45\textwidth]{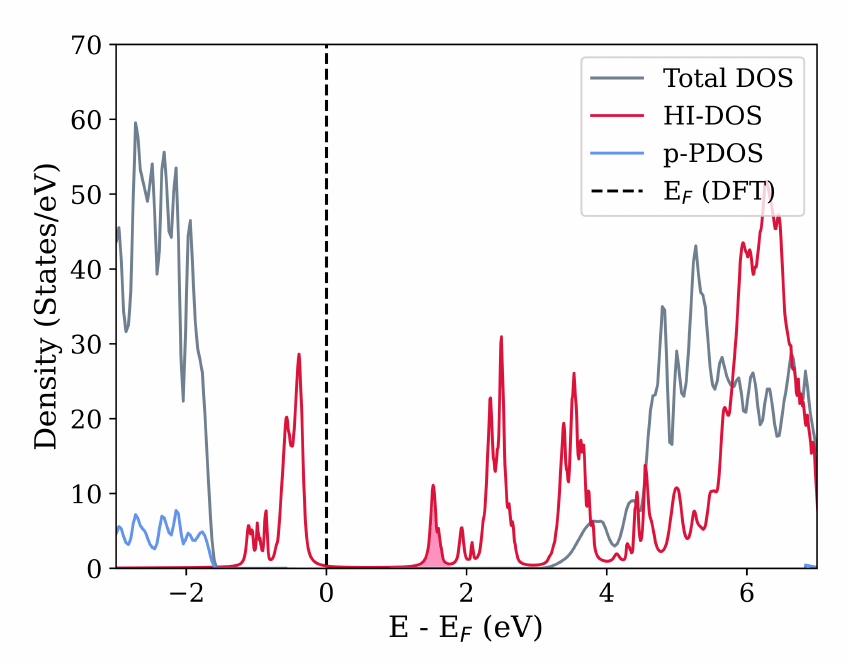}
    \caption{Combined the DFT and DFT+HI DOS. The hole-polaron peak is highlighted in light-red.}
    \label{fig:dos3}
\end{figure}

\subsection{Intersite Exchange Interactions in J$_{\mathrm{eff}}$ basis}

\begin{figure}[!h]
    \centering
    \includegraphics[width=0.9\textwidth]{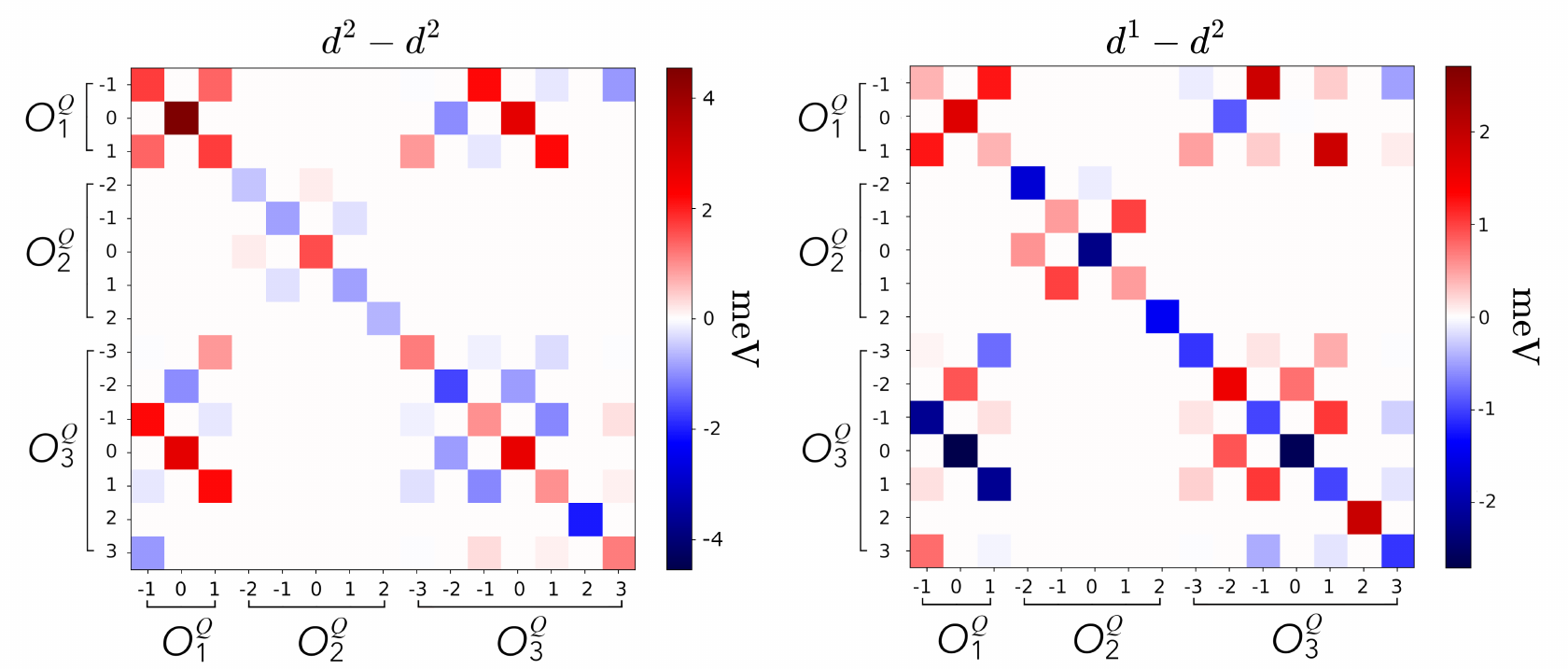}
    \caption{Color map of the IEI $V_{KK'}^{QQ'}$ in \bcnoo at pressure [0,1/2,0] Os-Os pair for (left) $d^2 -d^2$ interaction and (right) $d^1 -d^2$ interaction. All values are in meV. The numerical list of $V_{KK'}^{QQ'}$ is given in Suppl. Table~\ref{Tab:IEI3}.}
    \label{fig:iei3}
\end{figure}

In Suppl. Fig.~\ref{fig:iei3}, the IEI in \bcnoo are shown as a color map, with the corresponding values listed in Suppl. Table~\ref{Tab:IEI3} for the two cases of $d^2$–$d^2$ and $d^1$–$d^2$ electronic configurations. For the $d^1$–$d^2$ bond, an even stronger antisymmetric IEI matrix appears both in  quadrupole–quadrupole and dipole–octupole interactions induced by the polaron.

  	\renewcommand\floatpagefraction{0.1}
  \begin{table}[!h]
  	\caption{\label{Tab:IEI3}  
  		 Calculated IEI $V_{KK’}^{QQ’}$ for the $J_{\text{eff}}=3/2$/$J_{\text{eff}}=2$ multiplets. The first two columns list $Q$ and $Q’$, while the third and fourth columns display the $KQ$ and $K’Q’$ tensors in Cartesian representation. The last three columns report the IEI values (in meV) for the [0, 1/2, 0] nearest-neighbor Os–Os bond in the supercell reference frame, comparing: (i) $d^2$–$d^2$ interactions from Ref.~\cite{Pourovskii2021}, (ii) $d^2$–$d^2$ interactions from this work, and (iii) $d^1$–$d^2$ interactions from this work. A slash denotes the reversed $Q/Q’$ case, highlighting the asymmetric contributions.
  	}
	\begin{center}
		\begin{ruledtabular}
			\renewcommand{\arraystretch}{1.2}
			\begin{tabular}{c c c c c c c}
\multicolumn{4}{c}{Bonds $\mathbf{R}_{ij}$ = [0.5, 0.5, 0] (primitive cell)} & $d^2 -d^2$ (Ref.~\cite{Pourovskii2021}) & $d^2 - d^2$ & $d^1 - d^2$ \\
		\hline
\multicolumn{7}{c}{Dipole-Dipole} \\
 				\hline
-1 & -1   & y  & y  &   1.62  &    1.73  &    0.40  \\
 0 &  0   & z  & z  &   4.17  &    4.55  &    1.68  \\
 1 & -1   & x  & y  &   1.26  &    1.38  &    1.24  \\
 1 &  1   & x  & x  &   1.62  &    1.73  &    0.40  \\
		\hline
		\hline
\multicolumn{7}{c}{Quadrupole-Quadrupole} \\
\hline
-2 & -2   & xy  & xy                    &   -0.41  &   -0.52  &   -1.65  \\
-1 & -1   & yz  & yz                    &   -0.79  &   -0.84  &    0.52 \\
 0 & -2   & z$^2$  & xy                 &    0.16  &    0.15  &    0.56/-0.09  \\
 0 &  0   & z$^2$  & z$^2$              &    1.32  &    1.59  &   -2.25  \\
 1 & -1   & xz  & yz                    &   -0.23  &   -0.26  &    1.01  \\
 1 &  1   & xz  & xz                    &   -0.79  &   -0.84  &    0.52  \\
 2 &  2   & x$^2$-y$^2$  & x$^2$-y$^2$  &   -0.58  &   -0.67  &   -1.42  \\
\hline
		\hline
\multicolumn{7}{c}{Octupole-Octupole} \\
\hline
-3 & -3   & y(x$^2$-3y$^2$)  & y(x$^2$-3y$^2$)  &    1.16  &    1.15  &   -1.07  \\
-2 & -2   & xyz  & xyz                          &   -1.49  &   -1.64  &    1.51  \\
-1 & -3   & yz$^2$  & y(x$^2$-3y$^2$)           &   -0.14  &   -0.13  &    0.15  \\
-1 & -1   & yz$^2$  & yz$^2$                    &    0.80  &    0.96  &   -0.98  \\
 0 & -2   & z$^3$  & xyz                        &   -0.79  &   -0.87  &    0.74  \\
 0 &  0   & z$^3$  & z$^3$                      &    2.35  &    2.66  &   -2.61  \\
 1 & -3   & xz$^2$  & y(x$^2$-3y$^2$)           &   -0.29  &   -0.29  &    0.25/0.44  \\
 1 & -1   & xz$^2$  & yz$^2$                    &   -0.98  &   -1.05  &    1.04  \\
 1 &  1   & xz$^2$  & xz$^2$                    &    0.80  &    0.96  &   -0.98  \\
 2 &  2   & z(x$^2$-y$^2$)  & z(x$^2$-y$^2$)    &   -1.89  &   -2.04  &    1.92  \\
 3 & -1   & x(3x$^2$-y$^2$)  & yz$^2$           &    0.29  &    0.28  &   -0.25  \\
 3 &  1   & x(3x$^2$-y$^2$)  & xz$^2$           &    0.14  &    0.13  &   -0.14  \\
 3 &  3   & x(3x$^2$-y$^2$)  & x(3x$^2$-y$^2$)  &    1.16  &    1.15  &   -1.07  \\
		\hline
		\hline
\multicolumn{7}{c}{Dipole-Octupole} \\
\hline
-1 & -3   & y  & y(x$^2$-3y$^2$)    &          &   -0.02  &   -0.10/0.06  \\
-1 & -1   & y  & yz$^2$             &    1.97  &    2.20  &    1.87/-2.17  \\
-1 &  1   & y  & xz$^2$             &   -0.20  &   -0.20  &    0.26/0.17  \\
-1 &  3   & y  & x(3x$^2$-y$^2$)    &   -0.89  &   -0.92  &   -0.50/0.78  \\
 0 & -2   & z  & xyz                &   -0.97  &   -0.99  &   -0.88/0.89  \\
 0 &  0   & z  & z$^3$              &    2.38  &    2.71  &   -0.01/-2.71  \\
 0 &  2   & z  & z(x$^2$-3y$^2$)    &          &          &               \\
 1 & -3   & x  & y(x$^2$-3y$^2$)    &    0.89  &    0.92  &    0.50/-0.78  \\
 1 & -1   & x  & yz$^2$             &   -0.20  &   -0.20  &    0.26/0.17  \\
 1 &  1   & x  & xz$^2$             &    1.97  &    2.21  &    1.87/-2.17  \\
 1 &  3   & x  & x(3x$^2$-y$^2$)    &          &    0.02  &    0.10/-0.06  \\
  			\end{tabular}
\end{ruledtabular}
\end{center}
\end{table}

\clearpage
\subsection{Wavefunctions and Crystal Fields}

\begin{figure}[!h]
    \centering
    \includegraphics[width=0.9\textwidth]{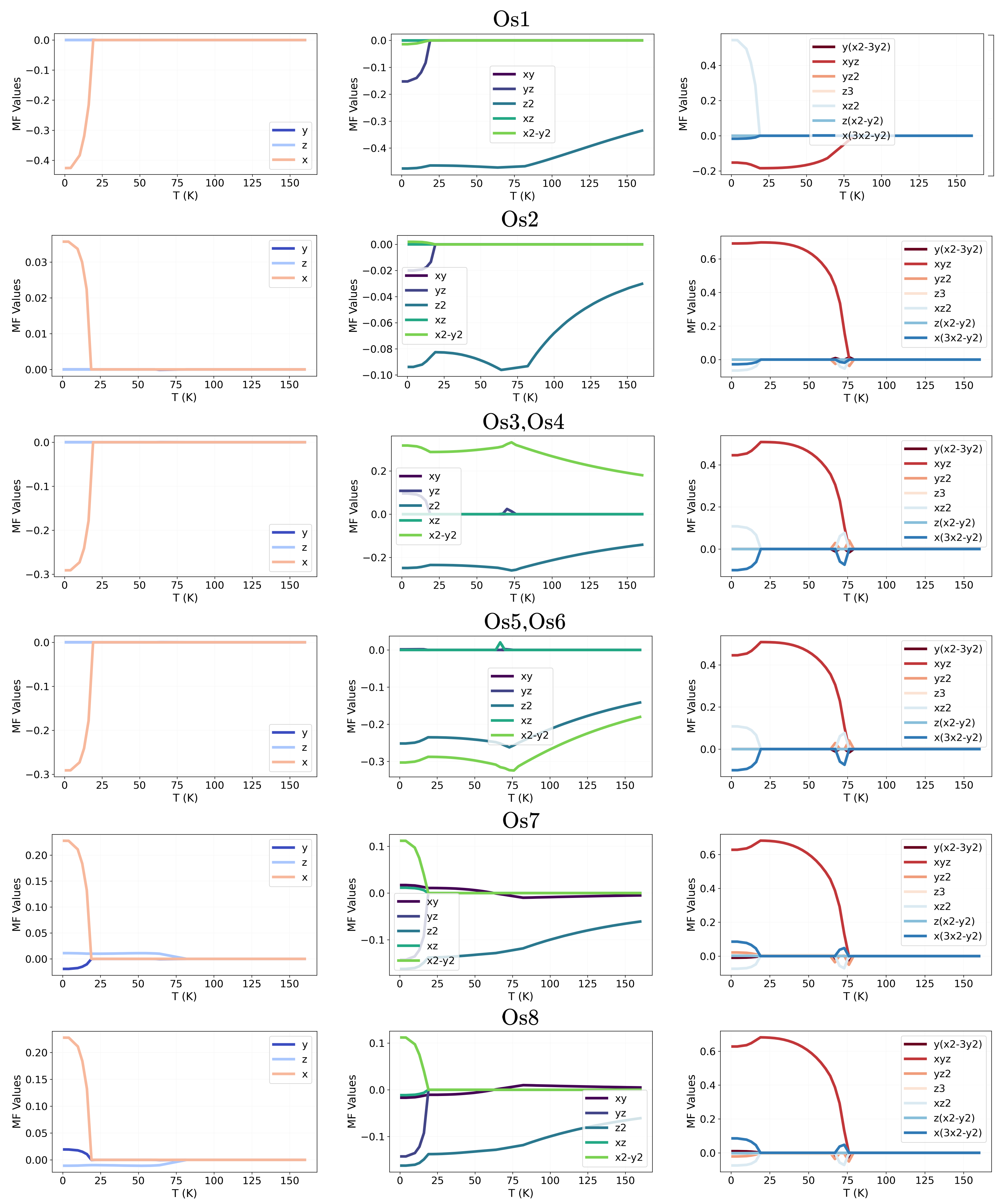}
    \caption{MF values of multipolar operators $O_K^{Q}$ expressed in Cartesian coordinates as a function of temperature for the different osmium ions of Fig.~\ref{fig:struct3}.}
    \label{fig:mf3}
\end{figure}

In the following, we report the crystal-field splittings and wavefunctions of the ground-state multiplets, expressed in the total angular momentum $J=3/2$ and $J=2$ basis. As in Sec. III,  the tetragonal symmetry of the chosen supercell and the broken symmetry brought by the polaron localization bring an equivalence between in-plane and out-of-plane osmium sites respectively. This feature manifests subsequently  in the MF results (see Fig.~\ref{fig:mf3}).

\vspace{1cm}

\noindent
\begin{minipage}[t]{0.48\linewidth}
\small\setlength{\tabcolsep}{6pt}\renewcommand{\arraystretch}{1.1}

\begin{tabular}{@{}lcccccc@{}}
\multicolumn{6}{c}{Energy (meV) for Os-1} \\
\hline
  & 0 & 20.3 &  &  &  \\
\hline
$\lvert 3/2,-3/2\rangle$ & $1/\sqrt{2}$    &  &       &       &  \\
$\lvert 3/2,-1/2\rangle$ &     &   $1/\sqrt{2}$    &  &    &       \\
$\lvert 3/3, 1/2\rangle$ &     &   $1/\sqrt{2}$    &       &       &       \\
$\lvert 3/2, 3/2\rangle$ &  $1/\sqrt{2}$   &       &  &  &       \\
\hline
\end{tabular}\\[0.8em]

\begin{tabular}{@{}lcccccc@{}}
\multicolumn{6}{c}{Energy (meV) for Os-2} \\
\hline
   & 0 & 0.9 & 17.4 & 17.5 & 17.5 \\
\hline
$\lvert 2,-2\rangle$ &     & $1/\sqrt{2}$ &  $1/\sqrt{2}$     &       &  \\
$\lvert 2,-1\rangle$ &     &       &  &    &   1    \\
$\lvert 2, 0\rangle$ & 1   &       &       &       &       \\
$\lvert 2, 1\rangle$ &     &       &  & 1 &       \\
$\lvert 2, 2\rangle$ &     & $1/\sqrt{2}$ &   $1/\sqrt{2}$    &       & \\
\hline
\end{tabular}\\[0.8em]

\begin{tabular}{@{}lcccccc@{}}
\multicolumn{6}{c}{Energy (meV) for Os-3} \\
\hline
   & 0 & 14.8 & 15.2 & 26.4 & 28.8 \\
\hline
$\lvert 2,-2\rangle$ & 0.323    & 0.629 &       &       & $1/\sqrt{2}$ \\
$\lvert 2,-1\rangle$ &     &       & $1/\sqrt{2}$ & $1/\sqrt{2}$&       \\
$\lvert 2, 0\rangle$ & 0.889   &   -0.457    &       &       &       \\
$\lvert 2, 1\rangle$ &     &       & $1/\sqrt{2}$ & -$1/\sqrt{2}$ &       \\
$\lvert 2, 2\rangle$ & 0.323   & 0.629 &       &       & -$1/\sqrt{2}$\\
\hline
\end{tabular}\\[0.8em]

\begin{tabular}{@{}lcccccc@{}}
\multicolumn{6}{c}{Energy (meV) for Os-4} \\
\hline
   & 0 & 14.8 & 15.2 & 26.4 & 28.8 \\
\hline
$\lvert 2,-2\rangle$ & 0.323    & 0.629 &       &       & $1/\sqrt{2}$ \\
$\lvert 2,-1\rangle$ &     &       & $1/\sqrt{2}$ & $1/\sqrt{2}$&       \\
$\lvert 2, 0\rangle$ & 0.889   &   -0.457    &       &       &       \\
$\lvert 2, 1\rangle$ &     &       & $1/\sqrt{2}$ & -$1/\sqrt{2}$ &       \\
$\lvert 2, 2\rangle$ & 0.323   & 0.629 &       &       & -$1/\sqrt{2}$\\
\hline
\end{tabular}\\[0.8em]

\end{minipage}\hfill
\begin{minipage}[t]{0.48\linewidth}
\small\setlength{\tabcolsep}{6pt}\renewcommand{\arraystretch}{1.1}

\begin{tabular}{@{}lcccccc@{}}
\multicolumn{6}{c}{Energy (meV) for Os-5} \\
\hline
   & 0 & 14.8 & 15.2 & 26.4 & 28.8 \\
\hline
$\lvert 2,-2\rangle$ & 0.323    & 0.629 &       &       & $1/\sqrt{2}$ \\
$\lvert 2,-1\rangle$ &     &       & $1/\sqrt{2}$ & $1/\sqrt{2}$&       \\
$\lvert 2, 0\rangle$ & 0.889   &   -0.457    &       &       &       \\
$\lvert 2, 1\rangle$ &     &       & $1/\sqrt{2}$ & -$1/\sqrt{2}$ &       \\
$\lvert 2, 2\rangle$ & 0.323   & 0.629 &       &       & -$1/\sqrt{2}$\\
\hline
\end{tabular}\\[0.8em]

\begin{tabular}{@{}lcccccc@{}}
\multicolumn{6}{c}{Energy (meV) for Os-6} \\
\hline
   & 0 & 14.8 & 15.2 & 26.4 & 28.8 \\
\hline
$\lvert 2,-2\rangle$ & 0.323    & 0.629 &       &       & $1/\sqrt{2}$ \\
$\lvert 2,-1\rangle$ &     &       & $1/\sqrt{2}$ & $1/\sqrt{2}$&       \\
$\lvert 2, 0\rangle$ & 0.889   &   -0.457    &       &       &       \\
$\lvert 2, 1\rangle$ &     &       & $1/\sqrt{2}$ & -$1/\sqrt{2}$ &       \\
$\lvert 2, 2\rangle$ & 0.323   & 0.629 &       &       & -$1/\sqrt{2}$\\
\hline
\end{tabular}\\[0.8em]

\begin{tabular}{@{}lcccccc@{}}
\multicolumn{6}{c}{Energy (meV) for Os-7} \\
\hline
   & 0 & 3.6 & 18.2 & 18.2 & 20.3 \\
\hline
$\lvert 2,-2\rangle$ &    & $1/\sqrt{2}$ &       &       & $1/\sqrt{2}$ \\
$\lvert 2,-1\rangle$ &     &       &  & 1  &       \\
$\lvert 2, 0\rangle$ &  1  &       &       &       &       \\
$\lvert 2, 1\rangle$ &     &       & 1 &  &       \\
$\lvert 2, 2\rangle$ &    & $1/\sqrt{2}$ &       &       & -$1/\sqrt{2}$\\
\hline
\end{tabular}\\[0.8em]

\begin{tabular}{@{}lcccccc@{}}
\multicolumn{6}{c}{Energy (meV) for Os-7} \\
\hline
   & 0 & 3.6 & 18.2 & 18.2 & 20.3 \\
\hline
$\lvert 2,-2\rangle$ &    & $1/\sqrt{2}$ &       &       & $1/\sqrt{2}$ \\
$\lvert 2,-1\rangle$ &     &       &  & 1  &       \\
$\lvert 2, 0\rangle$ &  1  &       &       &       &       \\
$\lvert 2, 1\rangle$ &     &       & 1 &  &       \\
$\lvert 2, 2\rangle$ &    & $1/\sqrt{2}$ &       &       & -$1/\sqrt{2}$\\
\hline
\end{tabular}\\[0.8em]

\end{minipage}

\subsection{Ordered Phase and Mean Field values with temperature}

In Fig.~\ref{fig:mf1}, we show the MF values of different multipolar operators in Cartesian coordinates. The sign reversal of the dominant octupolar exchange interaction remains present, although renormalized, and occurs at $T_N \approx 75$ K. At higher temperatures, only the quadrupoles of $z^2$ and $x^2-y^2$ character acquire finite expectation values, consistent with the previous cases (See Fig.~\ref{fig:mf1}). However, in this chemically doped configuration a second transition is observed at lower temperature ($\sim 25$ K), involving the simultaneous ordering of dipoles, quadrupoles, and octupoles. This transition is likely an artifact of the missing structural relaxation around the Na dopant, which is expected to significantly alter the local oxygen environment compared to the virtual-doping case discussed earlier. The spurious curves occurring around $T = 75$ K are a consequence of the poor MF convergence. 

\clearpage
\section{Constrained DFT+U ODMs}

In the following we report the ODMs used in the constrained DFT+U in the global reference frame of the supercell used, thus rotated by $45^o$ with respect to the local octahedral reference frame. 

\subsection{Polaron ($d^1$)}

\scriptsize
\begin{verbatim}
spin component  1

 0.61788  0.00000 -0.00000  0.00000  0.00000       -0.00000  0.00000 -0.00158  0.00000  0.01615
 0.00000  0.70678  0.00000  0.00001  0.00000        0.00000  0.00000  0.00000  0.08251  0.00000
-0.00000  0.00000  0.61789  0.00000 -0.01619        0.00158  0.00000 -0.00000  0.00000  0.00000
 0.00000  0.00001  0.00000  0.41890  0.00000        0.00000 -0.08251  0.00000 -0.00000  0.00000
-0.00000  0.00000 -0.01619  0.00000  0.56195       -0.01615  0.00000  0.00000  0.00000  0.00000

spin component  2

 0.00000  0.00589  0.00000  0.00000  0.00000        0.00000 -0.00000  0.00000 -0.02213  0.00000
 0.02213  0.00000  0.00000  0.00000 -0.00001       -0.00000  0.00000 -0.02219  0.00000 -0.22612
 0.00000 -0.00000  0.00000  0.02219  0.00000        0.00000  0.00591  0.00000 -0.00000  0.00000
-0.00000  0.00000 -0.00591  0.00000 -0.06021       -0.00589  0.00000 -0.00000  0.00000  0.00000
 0.00000  0.00000  0.00000 -0.22612  0.00000        0.00000 -0.06021  0.00000 -0.00001  0.00000

spin component  3

 0.00000  0.02213  0.00000  0.00000  0.00000        0.00000 -0.00000  0.00000  0.00589  0.00000
 0.00589  0.00000 -0.00000  0.00000  0.00000       -0.00000  0.00000 -0.00591  0.00000  0.06021
 0.00000 -0.00000  0.00000 -0.00591  0.00000        0.00000  0.02219  0.00000 -0.00000  0.00000
 0.00000  0.00000  0.02219  0.00000 -0.22612        0.02213  0.00000 -0.00000  0.00000  0.00001
 0.00000 -0.00001  0.00000 -0.06021  0.00000        0.00000  0.22612  0.00000 -0.00000  0.00000

spin component  4

 0.61788  0.00000  0.00000  0.00000  0.00000        0.00000  0.00000 -0.00158  0.00000 -0.01615
 0.00000  0.41890  0.00000  0.00001  0.00000        0.00000 -0.00000  0.00000 -0.08251  0.00000
 0.00000  0.00000  0.61789  0.00000  0.01619        0.00158  0.00000  0.00000  0.00000  0.00000
 0.00000  0.00001  0.00000  0.70678  0.00000        0.00000  0.08251  0.00000 -0.00000  0.00000
-0.00000  0.00000  0.01619  0.00000  0.56195        0.01615  0.00000  0.00000  0.00000 -0.00000
\end{verbatim}
\normalsize
\subsection{$d^2$}

\scriptsize
\centering
\begin{verbatim}
spin component  1

 0.57623  0.00000 -0.00000  0.00000 -0.00000        0.00000  0.00000 -0.00566  0.00000  0.03559
 0.00000  0.82697  0.00000 -0.00016  0.00000        0.00000 -0.00000  0.00000  0.12984  0.00000
 0.00000  0.00000  0.57630  0.00000 -0.04117        0.00566  0.00000 -0.00000  0.00000 -0.00002
 0.00000 -0.00016  0.00000  0.36763  0.00000        0.00000 -0.12984  0.00000  0.00000  0.00000
 0.00000  0.00000 -0.04117  0.00000  0.59576       -0.03559  0.00000  0.00002  0.00000  0.00000
 
spin component  2

 0.00000  0.01762  0.00000 -0.00000  0.00000        0.00000 -0.00000  0.00000 -0.05326  0.00000
 0.05326  0.00000  0.00002  0.00000  0.00013       -0.00000  0.00000 -0.05180  0.00000 -0.35948
 0.00000 -0.00002  0.00000  0.05180  0.00000        0.00000  0.01050  0.00000 -0.00002  0.00000
 0.00000  0.00000 -0.01050  0.00000 -0.09904       -0.01762  0.00000  0.00002  0.00000 -0.00013
 0.00000 -0.00013  0.00000 -0.35948  0.00000        0.00000 -0.09904  0.00000  0.00013  0.00000

spin component  3

 0.00000  0.05326  0.00000 -0.00000  0.00000        0.00000  0.00000  0.00000  0.01762  0.00000
 0.01762  0.00000 -0.00002  0.00000 -0.00013       -0.00000  0.00000 -0.01050  0.00000  0.09904
 0.00000  0.00002  0.00000 -0.01050  0.00000        0.00000  0.05180  0.00000 -0.00002  0.00000
 0.00000  0.00000  0.05180  0.00000 -0.35948        0.05326  0.00000  0.00002  0.00000 -0.00013
 0.00000  0.00013  0.00000 -0.09904  0.00000        0.00000  0.35948  0.00000  0.00013  0.00000

spin component  4

 0.57623  0.00000  0.00000  0.00000 -0.00000       -0.00000  0.00000 -0.00566  0.00000 -0.03559
 0.00000  0.36763  0.00000 -0.00016  0.00000        0.00000  0.00000  0.00000 -0.12984  0.00000
-0.00000  0.00000  0.57630  0.00000  0.04117        0.00566  0.00000  0.00000  0.00000  0.00002
 0.00000 -0.00016  0.00000  0.82697  0.00000        0.00000  0.12984  0.00000 -0.00000  0.00000
 0.00000  0.00000  0.04117  0.00000  0.59576        0.03559  0.00000 -0.00002  0.00000 -0.00000
\end{verbatim}
\normalsize

\bibliography{bibliography}